\documentclass[acmtog,authorversion]{acmart}
\usepackage{booktabs} 

\usepackage[ruled,linesnumbered]{algorithm2e}

\setcopyright{acmcopyright}
\acmJournal{TOG}
\acmYear{2018}
\acmVolume{37}
\acmNumber{4}
\acmArticle{42}
\acmMonth{8} 
\acmDOI{10.1145/3197517.3201290}

\setcitestyle{square}

\usepackage{amssymb}
\usepackage{bm}
\usepackage{textcomp}
\usepackage{multirow}
\usepackage{bigstrut}
\usepackage{array}
\usepackage{gensymb}
\usepackage{pifont}
\usepackage{subfigure}
\usepackage{graphicx}
\usepackage{psfrag}
\usepackage{pstricks}
\usepackage{microtype}
\usepackage[mathscr]{euscript}
\usepackage{enumitem}

\newcommand{\eigen}{\textsc{Eigen}}
\newcommand{\libigl}{\textsc{Libigl}}
\newcommand{\para}[1]{\subsubsection*{#1}}

\DeclareMathOperator*{\argmin}{arg\,min}
\newcommand{\localstep}{\textit{Local-Step}}
\newcommand{\globalstep}{\textit{Global-Step}}
\begin{document}
\title{Anderson Acceleration for Geometry Optimization and Physics Simulation} 

\author{Yue Peng}
\affiliation{%
  \institution{University of Science and Technology of China}
}
\email{echoyue@mail.ustc.edu.cn}

\author{Bailin Deng}
\affiliation{%
  \institution{Cardiff University}
}
  \email{DengB3@cardiff.ac.uk}

\author{Juyong Zhang}
\affiliation{%
  \institution{University of Science and Technology of China}
}
  \email{juyong@ustc.edu.cn}
\authornote{Corresponding author (\href{mailto:juyong@ustc.edu.cn}{juyong@ustc.edu.cn}).}

\author{Fanyu Geng}
\affiliation{%
  \institution{University of Science and Technology of China}
}
\email{gfy29110@mail.ustc.edu.cn}

\author{Wenjie Qin}
\affiliation{%
  \institution{University of Science and Technology of China}
}
\email{darkqin@mail.ustc.edu.cn}

\author{Ligang Liu}
\affiliation{%
  \institution{University of Science and Technology of China}
}
\email{lgliu@ustc.edu.cn}

\authorsaddresses{Authors' addresses:
$\{$Yue Peng, Juyong Zhang, Fanyu Geng, Wenjie Qin, Ligang Liu$\}$, University of Science and Technology of China, 96 Jinzhai Road, Hefei, Anhui, 230026, China; $\{$echoyue@mail.ustc.edu.cn, juyong@ustc.edu.cn, gfy29110@mail.ustc.edu.cn, \mbox{darkqin@mail.ustc.edu.cn}, lgliu@ustc.edu.cn$\}$;
Bailin Deng, Cardiff University, 5 The Parade, Cardiff CF24 3AA, Wales, United Kingdom, DengB3@cardiff.ac.uk}

\renewcommand{\shortauthors}{Peng, Deng, Zhang, Geng, Qin, and Liu}

\begin{abstract}
Many computer graphics problems require computing geometric shapes subject to certain constraints. This often results in non-linear and non-convex optimization problems with globally coupled variables, which pose great challenge for interactive applications. Local-global solvers developed in recent years can quickly compute an approximate solution to such problems, making them an attractive choice for applications that prioritize efficiency over accuracy. However, these solvers suffer from lower convergence rate, and may take a long time to compute an accurate result. In this paper, we propose a simple and effective technique to accelerate the convergence of such solvers. By treating each local-global step as a fixed-point iteration, we apply Anderson acceleration, a well-established technique for fixed-point solvers, to speed up the convergence of a local-global solver. To address the stability issue of classical Anderson acceleration, we propose a simple strategy to guarantee the decrease of target energy and ensure its global convergence. In addition, we analyze the connection between Anderson acceleration and quasi-Newton methods, and show that the canonical choice of its mixing parameter is suitable for accelerating local-global solvers. Moreover, our technique is effective beyond classical local-global solvers, and can be applied to iterative methods with a common structure. We evaluate the performance of our technique on a variety of geometry optimization and physics simulation problems. Our approach significantly reduces the number of iterations required to compute an accurate result, with only a slight increase of computational cost per iteration. Its simplicity and effectiveness makes it a promising tool for accelerating existing algorithms as well as designing efficient new algorithms.
\end{abstract}

%
%


%
%


\begin{teaserfigure}
	\centering
	\includegraphics[width=\textwidth]{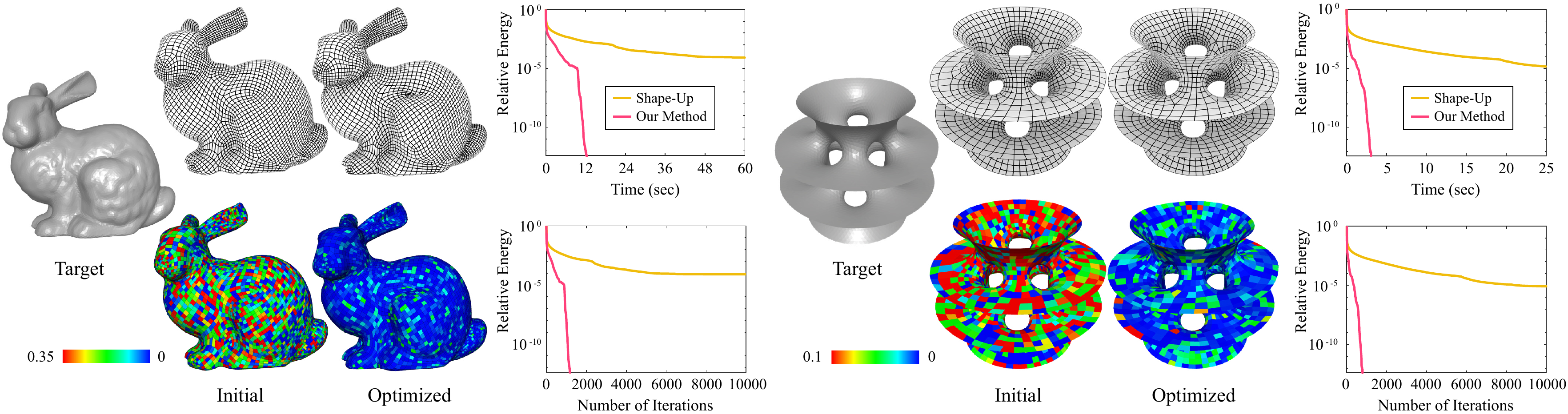}
	\caption{Compared with the local-global solver from~\cite{BouazizDSWP12}, our method significantly reduces the computational time and iterations required for an accurate solution to planar quad mesh optimization, as shown in the log-scale relive energy error graphs. The color coding shows the planarity error for each face, computed as the max distance from its vertices and their best fitting plane, normalized by the average edge length of the mesh.}
	\label{Fig:Planarity}
\end{teaserfigure}

\maketitle

\section{Introduction}

Many computer graphics problems require computing geometric shapes whose elements are subject to certain constraints. In geometric modeling, such constraints can be related to aesthetics, performance, or fabrication requirements of the shape. For example, for freeform architectural design represented as quadrilateral meshes, it is often desirable that each face is planar such that the panel can be easily constructed~\cite{Liu2006}. In physics simulation, geometric constraints can be used for defining potential energies that determine the deformation behavior of an object. For example, the elastic potential energy of a spring can be defined via a constraint on its length~\cite{Liu2013}. In this paper, we focus on shapes that are represented using a discrete set of points, such as the vertices of a mesh or the nodes of a physical system. When dealing with such shapes subject to geometric constraints, a common task is to determine the shape that best satisfies the constraints, by solving an optimization problem about its point positions $\mathbf{Q}$
\begin{equation}
	\min_{\mathbf{Q}} = \sum_i E_i(\mathbf{Q}),
	\label{eq:OptProblem}
\end{equation}
where functions $E_i$ measures the violation of the constraints. For many computer graphics problems, the constraints are non-linear, and each point among $\mathbf{Q}$ is involved in multiple constraints. Consequently, the optimization problem is often non-convex, with globally coupled variables. For interactive applications such as shape exploration and physics simulation, the optimization needs to be done repeatedly according to the user input or the current state of the physical system, and the solution needs to be computed efficiently in order to provide real-time feedback for the user. Such requirement of efficiency poses great challenges to traditional Newton-type solvers, because in each iteration these solvers need to evaluate the gradient and the Hessian of the target function, and solve a linear system to update the variables; both steps can be time-consuming, especially for large-scale problems~\cite{nocedal2006numerical}. 

In recent years, first-order methods have become an increasingly popular choice for solving large-scale optimization problems. These methods utilize information on the value or gradient of the target function but not the Hessian, to reduce the computational cost per iteration~\cite{Beck2017}. Within computer graphics, first-order methods have been developed for optimization problems in the form of~\eqref{eq:OptProblem} that arise in geometry processing~\cite{BouazizDSWP12} and physics simulation~\cite{Bouaziz2014}. The main idea of~\cite{BouazizDSWP12} and~\cite{Bouaziz2014} is to reformulate the target function using auxiliary variables that represent the projections of point positions onto the feasible sets of geometric constraints, and to minimize the target function in a local-global manner: the local step fixes the point positions and updates the auxiliary variables, and reduces to evaluating closed-form projection operators for the constraints; the global step fixes the auxiliary variables and updates the point positions, which amounts to solving a sparse positive definite linear system with a fixed matrix. The main strength of such local-global solvers is their efficiency and robustness. Both the local and the global steps have a simple form that can be efficiently solved and allows for parallelization, and guarantee decrease of the target energy unless a local minimum has been reached. Moreover, the solver rapidly decreases the target energy within the initial iterations, and quickly produces an approximate solution. These properties make them well suited for interactive applications where efficiency and stability are prioritized.

On the other hand, the efficiency of local-global solvers comes at the cost of accuracy. Although they are efficient to produce an approximate solution, it can take a very long time to compute an accurate result~\cite{Grinspun14b}, because their convergence rate is sublinear in general~\cite{Beck2013}. In this paper, we propose a simple method to accelerate the convergence of these solvers, while retaining the computational efficiency that makes them attractive for interactive applications. Our key idea is to treat the sequence generated by these solvers as fixed-point iteration, and to speed up their convergence using Anderson acceleration, a well-established technique for fixed-point solvers~\cite{Walker2011}. 
Originally proposed for solving nonlinear integral equations~\cite{Anderson1965}, Anderson acceleration has achieved great success in computational chemistry~\cite{Pulay1980,Pulay1982} and become a standard procedure for accelerating electronic structure calculation algorithms~\cite{Rohwedder2011}. The past few years have seen revived research interest from the numerical analysis community on the theory and applications of Anderson acceleration~\cite{Fang2009,Walker2011,Lipnikov2013,Toth2015,Higham2016,Toth2017}, 
as well as its growing applications in other domains such as computational physics~\cite{Willert2014,Pratapa2016,An2017}. Within the computer graphics community, Anderson acceleration has been unknown and unexplored. In this paper, we show its effectiveness on accelerating local-global solvers in geometry optimization and physics simulation, as well as other solvers that share a similar structure. 

Although Anderson acceleration has been successful for a variety of problems in different domains, there are a few known challenges that we must overcome in order to apply it to the local-global solvers. First, despite a recent proof of its local convergence, Anderson acceleration lacks guarantee of global convergence, and may stagnate when started from a point far from the solution~\cite{Potra2013}.
To improve its robustness, within each iteration we revert to the local-global iterate if the accelerated iterate increase the target energy. Such strategy guarantees monotonic decrease of the energy, with only slight increase of computational cost per iteration. 

Another challenge is that Anderson acceleration relies on a mixing parameter $\beta$ which can impact its performance. Its optimal value is problem dependent~\cite{Fang2009}, and existing works simply set this parameter to an empirical value.
We analyze the connection between Anderson acceleration and quasi-Newton methods, and show that the canonical choice $\beta = 1$ is suitable for the problems considered in this paper. In particular, we show that Anderson acceleration is closely related to the L-BFGS method proposed in~\cite{LiuBK17} for accelerating physics simulation: in each iteration, both methods start with an equivalent choice of approximate Hessian, but use different strategies of adapting it to a limited amount of previous iterations.

We test the performance of our algorithm on a variety of geometry processing and physics simulation problems. Our experiments show that it significantly reduces the number of iterations required to compute a high-accuracy solution, with only slightly increased computational cost per iteration compared to the local-global solvers. As a result, the technique can be applied to increase the solution accuracy within the same computational time budget, or to reduce the computational cost for a solution with the same accuracy. 

Beyond local-global solvers, Anderson acceleration is applicable for other iterative solvers, as long as each iteration deceases the target energy and two consecutive iterates are related by well-defined mappings. We propose a general procedure for applying Anderson acceleration to such solvers, and showcase its effectiveness on two geometric computing algorithms. Given the popularity of fixed-point iterations in computer graphics, Anderson acceleration can be a promising tool for speeding up existing algorithms, as well as designing efficient new algorithms.

To summarize, our main contributions include:
\begin{itemize}[leftmargin=*]
	\item We adapt Anderson acceleration to speed up the convergence of local-global solvers for geometry optimization and physics simulation. We propose a simple and effective strategy to guarantee the monotonic decrease and global convergence of the target energy, with low computational overhead. 
	\item We analyze the relation between Anderson acceleration and other quasi-Newton methods such as L-BFGS, and show the effectiveness of the canonical mixing parameter for our target problems.
	\item We analyze the effectiveness of Anderson acceleration beyond local-global solvers, and propose a general approach for applying it to the applicable solvers.
\end{itemize} 
\section{Related Work}

\para{Local-global solvers}
Due to the efficiency for computing an approximate solution, local-global solvers have been applied to various computer graphics problems from shape modeling to physics simulation. Sorkine and Alexa~\shortcite{SorkineA07} performed as-rigid-as-possible (ARAP) surface modeling by minimizing a target energy that measures local rigidity of the surface; using auxiliary variables to represent the closest rigid deformation for each local element, the energy is minimized in a local-global manner. Liu et al.~\shortcite{Liu2008} adapted this approach to compute conformal or isometric parameterization for triangle meshes in a least-squares manner. Bouaziz et al.~\shortcite{BouazizDSWP12} proposed a unified local-global optimization framework for geometric constraints, by formulating a target function that is the weighted sum of squared distance from the constrained elements to their feasible shapes, with auxiliary variables representing the closest feasible configurations. Liu et al.~\shortcite{Liu2013} adopted local-global formulation for implicit Euler integration of mass-spring simulation, resulting in much faster results that are visually close to the true solutions. Bouaziz et al.~\shortcite{Bouaziz2014} extended this approach to \emph{projective dynamics}, a general framework for implicit time integration of physical systems, by using geometric constraints to define potential energies.

Local-global solvers may take a long time to converge to a high-accuracy solution. Different techniques have been proposed recently to address this issue. Based on the similarity between projective dynamics and iterative linear system solving, Wang~\shortcite{Wang15} proposed to use the Cheybysheve semi-iterative method to speed up the convergence of projective dynamics. The technique was later improved in~\cite{Wang2016} to derive a preconditioned gradient descent method suitable for GPU implementation. Liu et al.~\shortcite{LiuBK17} applied L-BFGS to minimize the projective dynamics target energy, resulting in faster convergence than the local-global solver. 

\para{Fast geometry optimization}
Besides local-global solvers and their accelerated versions, there are other fast solvers for geometric optimization. Tang et al.~\shortcite{Tang2014} formulated various geometric constraints as quadratic equations, which are solved via non-linear least squares optimization using the Gauss-Newton algorithm. Kovalsky et al.~\shortcite{KovalskyGL16} accelerate the optimization of geometric energies by minimizing a local convex quadratic proxy function to achieve preconditioning effects, combined with an acceleration similar to Nesterov's acceleration method~\cite{Nesterov83}. Rabinovich et.al.~\shortcite{RabinovichPPS17} presented a scalable approach for optimizing flip-preventing energies, using a reweighted proxy function in each iteration. Shtengel et al.~\shortcite{ShtengelPSKL17} derived convex majorizers of composite energies via convex-concave decompositions, which are repeatedly updated and minimized to obtain a minimum of the target energy. 

\para{Anderson acceleration}
In the past, Anderson acceleration has been independently developed by different authors, and successfully applied in various problem domains for improving convergence of iterative solvers. It was originally proposed by D. G. Anderson in 1965 for iterative solution of nonlinear integral equations~\cite{Anderson1965}. Later, Pulay~\shortcite{Pulay1980,Pulay1982} introduced the same technique for stabilizing and accelerating the convergence of self-consistent field method in quantum chemistry computation. Pulay's method, often called \emph{direct inversion in the iterative subspace} (DIIS) or \emph{Pulay mixing}, proves to be effective for a much broader range of iterative solvers used in electronic structure calculations, and has been a standard practice for accelerating these algorithms~\cite{Rohwedder2011}. Anderson acceleration has also been proposed as a Krylov subspace acceleration technique, for solving nonlinear partial differential equations~\cite{Washio1997,Oosterlee2000}.
In recent years, there is emerging interest from the numerical analysis community in the theories and applications of Anderson acceleration. From a theoretical perspective, Anderson acceleration is shown to be a kind of quasi-Newton method for solving the nonlinear equations, which utilizes the previous $m$ iterates to approximate the inverse Jacobian~\cite{Eyert1996,Fang2009,Rohwedder2011}. Toth and Kelley~\shortcite{Toth2015} proved the local convergence of Anderson acceleration for contractive fixed-point iterations. The convergence result was later extended to fixed-point maps corrupted with errors~\cite{Toth2017}. In terms of applications, Walker and Ni~\shortcite{Walker2011} showcase its effectiveness for various numerical problems such statistical estimation, nonnegative matrix factorization, and domain decomposition. It is also used by Lipnikov et al.~\shortcite{Lipnikov2013} to accelerate the numerical solving of advection-diffusion problems. Other applications include low-rank tensor approximation~\cite{Sterck2012}, solving large sparse linear systems~\cite{Pratapa2016,Suryanarayana2016}, and fluid dynamics~\cite{Ho2017}, just to name a few.
\section{Background}
\label{sec:Background}
In this section, we review the local-global solvers for geometry optimization and physics simulation, to prepare for the discussion of their acceleration in Section~\ref{sec:AA}.

\subsection{Geometry optimization}
Given a shape represented with points $\mathbf{q}_1, \ldots, \mathbf{q}_n \in \mathbb{R}^d$ subject to a set of geometric constraints, Bouaziz et al.~\shortcite{BouazizDSWP12} compute the shape that best satisfies the constraints by optimizing the point positions
\begin{equation}
\min_{\mathbf{Q}, \{\mathbf{P}_i\}}
\sum_{i} \frac{w_i}{2} \| \mathbf{A}_i \mathbf{Q} - \mathbf{P}_i  \|_F^2 + \sigma_i(\mathbf{P}_i).
\label{eq:GeometryOpt}
\end{equation}
Here matrix $\mathbf{Q} \in \mathbb{R}^{n \times d}$ stacks all the point positions. Matrix $\mathbf{A}_i \in \mathbb{R}^{k_i \times n}$ selects the relevant points for constraint $i$ and applies linear transformation to derive an appropriate representation; for example, $\mathbf{A}_i$ can represent subtracting the mean position from the relevant points for a translation-invariant constraint, to achieve faster convergence for the solver~\cite{BouazizDSWP12}. $\mathbf{P}_i \in \mathbb{R}^{k_i \times d}$ are auxiliary variables representing the closest projection of $\mathbf{A}_i \mathbf{Q}$ onto the feasible set $\mathcal{C}_i$ of constraint $i$. Thus $\| \mathbf{A}_i \mathbf{Q} - \mathbf{P}_i  \|_F^2$ is a \emph{shape proximity function} that measures the distance from constrained elements to their closest feasible configuration, with a weight $w_i$ specified by the user to control its importance. $\sigma_i$ is an indicator function for feasible set $\mathcal{C}_i$ to ensure $\mathbf{P}_i$ satisfies the constraint:
\[
	\sigma_i(\mathbf{P}_i)
	= \left\{
	\begin{array}{ll}
		0 & \textrm{if}~\mathbf{P}_i \in \mathcal{C}_i,\\
		+\infty & \textrm{otherwise}.
	\end{array}
	\right.
\]
This formulation allows us to write other regularization energies in a unified way. For example, the Laplacian energy can be represented by setting $\mathbf{A}_i$ to a row of the Laplacian matrix, with $\mathcal{C}_i = \{\mathbf{0}\}$.

The optimization problem is solved using block coordinate descent, by alternating between two steps: 
\begin{itemize}[leftmargin=*]
	\item In the local step, the target energy is minimized with respect to $\{\mathbf{P}_i\}$ while fixing $\mathbf{Q}$. This reduces to independent sub-problems of projecting each $\mathbf{A}_i \mathbf{Q}$ onto feasible set $\mathcal{C}_i$, which can be solved in parallel. Moreover, for many geometric constraints, the projection operator has closed-form representation that can be efficiently evaluated~\cite{BouazizDSWP12}.
	\item In the global step, we fix $\{\mathbf{P}_i\}$ and minimize the energy with respect to $\mathbf{Q}$. This reduces to solving a sparse symmetric positive definite system with $d$ right-hand-sides:
	\begin{equation}
	\left( \sum_i w_i \mathbf{A_i^T} \mathbf{A}_i \right) \mathbf{Q} = \sum_i w_i \mathbf{A}_i^T \mathbf{P}_i.  
	\label{eq:GlobalStep}
	\end{equation}
	Since the system matrix is fixed for all iterations, we can pre-compute its Cholesky factorization and efficiently solve for each right-hand-sides in parallel.
\end{itemize}
Both steps are highly efficient with parallelization. In addition, each step is guaranteed to lower the target energy unless a local minimum has been reached, thus the solver is guaranteed to converge. 
Furthermore, within a small number of iterations, the solver rapidly decreases the target energy and produces an approximate solution. Such properties make it an attractive choice for applications where efficiency is prioritized over high accuracy, such as interactive constraint-based modeling. 

\subsection{Physics simulation}
The local-global solving strategy has also been applied for physics simulation~\cite{Liu2013,Bouaziz2014}. In particular, projective dynamics~\cite{Bouaziz2014} performs implicit time integration by solving an optimization problem similar to~\eqref{eq:GeometryOpt}. Given a physical system consisting of $n$ nodes with positions $\overline{\mathbf{Q}} \in \mathbb{R}^{n \times 3}$ and velocities $\overline{\mathbf{V}} \in \mathbb{R}^{n \times 3}$ at time instance $t$, the node positions $\mathbf{Q}$ at time $t + h$ are computed by solving 
\begin{equation}
\min_{\mathbf{Q}, \{\mathbf{P}_i\}} \frac{1}{2h^2}\|\mathbf{M}^{\frac{1}{2}}(\mathbf{Q} - \mathbf{R})\|_F^2 + \sum_i \frac{w_i}{2}\| \mathbf{A}_i \mathbf{Q} - \mathbf{P}_i \|_F^2 + \sigma_i(\mathbf{P}_i).
\label{eq:ProjDyna}
\end{equation}
Here $\mathbf{M} \in \mathbb{R}^{n \times n}$ is the mass matrix, and $\mathbf{R} = \overline{\mathbf{Q}} + h \overline{\mathbf{V}} + h^2 \mathbf{M}^{-1} \mathbf{F}_{\mathrm{ext}}$ is a momentum term where $\mathbf{F}_{\mathrm{ext}} \in \mathbb{R}^{n \times 3}$ stores the external forces. The remaining terms $\frac{w_i}{2}\| \mathbf{A}_i \mathbf{Q} - \mathbf{P}_i \|_F^2 + \sigma_i(\mathbf{P}_i)$ are the same as in Equation~\eqref{eq:GeometryOpt}, for measuring the squared distance to feasible sets of geometric constraints. With appropriate geometric constraints and weights, each term  $\frac{w_i}{2}\| \mathbf{A}_i \mathbf{S}_i \mathbf{Q} - \mathbf{P}_i \|_F^2$ defines a potential energy whose gradient equals to the induced internal forces for the relevant nodes. For example, for a spring between two nodes $\mathbf{q}_{i_1}, \mathbf{q}_{i_2}$, the potential energy is defined using length constraint $\|\mathbf{q}_i - \mathbf{q}_j\| = L$ where $L$ is the rest length of the spring, with the weight $w_i$ being the spring stiffness~\cite{Liu2013}. Similar to geometry optimization, the problem~\eqref{eq:ProjDyna} is solved via block coordinate descent, with the local step updating $\{\mathbf{P}_i\}$ via projection, and the global step updating $\mathbf{Q}$ by solving a sparse SPD linear system~\cite{Bouaziz2014}:
\begin{equation}
\left(\frac{1}{h^2} \mathbf{M} + \sum_i w_i \mathbf{A}_i^T \mathbf{A}_i  \right) \mathbf{Q} = \frac{1}{h^2} \mathbf{M} \mathbf{R} + \sum_i w_i \mathbf{A}_i^T \mathbf{P}_i.
\label{eq:Physics}
\end{equation}
Then the velocities at time $t+h$ are computed as $\mathbf{V} = (\mathbf{Q} - \overline{\mathbf{Q}})/h$.


\section{Our Method}
\label{sec:AA}

Despite the fast convergence of local-global solvers to an approximate solution, it can take a much longer time to produce an accurate solution. This is because in general the target function is nonconvex, and for such problems the convergence rate of block coordinate descent is only sublinear~\cite{Beck2013}. In the past, different approaches have been proposed to speed up the convergence of local-global solvers~\cite{Wang15,LiuBK17}. In this section, we present a new approach based on Anderson acceleration, analyze its performance, and compare it with existing approaches.

\subsection{Anderson acceleration}
To accelerate the convergence of the local-global solvers, we first note that in the local step, the updated projection $\mathbf{P}_i$ is a function of the current positions $\mathbf{Q}$. Therefore, the local step and the global step can be combined into a fixed-point iteration
\begin{equation}
	\mathbf{Q}_{\mathrm{LG}}^{k} = G(\mathbf{Q}^{k-1}).
	\label{eq:FixedPointFormat}
\end{equation}
For example, for the geometry optimization problem~\eqref{eq:GeometryOpt}, the mapping $G$ becomes
\[
	G(\cdot) = \left( \sum_i w_i \mathbf{A_i^T} \mathbf{A}_i \right)^{-1} \sum_i w_i\mathbf{A}_i^T \mathbf{P}_i(\cdot)
\]
This perspective enables us to apply Anderson Acceleration~\cite{Anderson1965}, a well-established technique for fixed-point iterations, to speed up the convergence. Note that for a solution $\mathbf{Q}^{\ast}$ to the fixed-point iteration~\eqref{eq:FixedPointFormat}, the \emph{residual}
\begin{equation}
F(\mathbf{Q}) = G(\mathbf{Q}) - \mathbf{Q}
\label{eq:Residual}
\end{equation}
must vanish. The key idea of Anderson Acceleration is to utilize the current iterate $\mathbf{Q}^k$ as well as the previous  $m$ iterates $\mathbf{Q}^{k-1}, \ldots, \mathbf{Q}^{k-m}$, to derive a new iterate $\mathbf{Q}_{\mathrm{AA}}^{k+1}$ that decreases the residual norm as much as possible.
Specifically, $\mathbf{Q}^{k}, \mathbf{Q}^{k-1}, \ldots, \mathbf{Q}^{k-m}$ span an affine subspace where each point can be written as
\[
\mathbf{Q}(\boldsymbol{\alpha}) = \mathbf{Q}^{k} + \sum_{j=1}^m \alpha_j (\mathbf{Q}^{k-j} - \mathbf{Q}^{k}),
\]
with $\boldsymbol{\alpha} = (\alpha_1, \ldots, \alpha_m)$ being its affine coordinates. Within this subspace, we derive an approximation $\widetilde{G}$ of the mapping $G$ via barycentric interpolation:
\begin{equation}
	\widetilde{G}\left(\mathbf{Q}(\boldsymbol{\alpha})\right) = G(\mathbf{Q}^{k}) + \sum_{j=1}^m \alpha_j \left(G(\mathbf{Q}^{k-j}) - G(\mathbf{Q}^{k})\right).
	\label{eq:BarycentricInterpolation}
\end{equation}
Using this model, we can find the point $Q(\boldsymbol{\alpha}^\ast)$ with the smallest residual norm, by solving a linear least-squares problem
\begin{align}
	\boldsymbol{\alpha}^{\ast} &= \argmin_{\boldsymbol{\alpha}} \left\| \widetilde{G}\left(\mathbf{Q}(\boldsymbol{\alpha})\right) - \mathbf{Q}(\boldsymbol{\alpha}) \right\|^2 \nonumber\\
	&= \argmin_{\boldsymbol{\alpha}} \left\| F^k + \sum_{j=1}^m \alpha_j (F^{k-j} - F^{k})\right\|^2,
	\label{eq:LS}
\end{align}
where $F^k = G(\mathbf{Q}^k) - \mathbf{Q}^k$ is the residual at iteration $k$. Then the new iterate is computed by combining $Q(\boldsymbol{\alpha}^\ast)$ and $\widetilde{G}\left(\mathbf{Q}(\boldsymbol{\alpha}^\ast)\right)$:
\begin{equation}
	\mathbf{Q}_{\mathrm{AA}}^{k+1} = (1 - \beta) Q(\boldsymbol{\alpha}^\ast) + \beta \widetilde{G}(\mathbf{Q}(\boldsymbol{\alpha}^\ast)),
	\label{eq:FullAAUpdate}
\end{equation}
where $\beta \in (0, 1]$ is a \emph{mixing parameter}. The majority of existing work simply choose $\beta = 1$, and we will follow this convention in the current paper. Later in Section~\ref{sec:Analysis} we will show that this is indeed a suitable choice for the considered problems. The value $m$ is typically no larger than 6 (see Sec.~\ref{sec:WindowSizeChoice} for discussion on the choice of $m$). By taking the previous $m$ iterates into account, the local model $\widetilde{G}$ captures the variation of $G$ around the current iterate, which results in better convergence behavior. Figure~\ref{Fig:strategy} shows an example, where Anderson acceleration significantly speed up the convergence of planar quad mesh optimization.

Anderson acceleration can be considered as a multi-dimensional generalization of the secant method for root-finding (\cite{Kelley1995}, Section 5.4.5). In each iteration, the secant method approximates a univariate function graph using the line between two points on the graph that are evaluated at the latest two iterates, and intersect, and finds the root of this line as the next iterate. Despite this seemingly poor approximation, the secant method achieves local super-linear convergence~\cite{Kelley1995}. 
Similarly, Anderson acceleration finds the root of a multivariate function $F(Q)=G(Q)-Q$, by iteratively approximating the graph $(Q,F(Q))$ using the affine space spanned by multiple points evaluated at the latest $m$ iterates, which is equivalent to the barycentric interpolation~\eqref{eq:BarycentricInterpolation}. For this reason, it is called a \emph{multisecant} method by some authors~\cite{Fang2009}.

Note that solving the problem~\eqref{eq:LS} requires recomputing the differences between $F^k$ and all $m$ previous residuals to update the whole least-squares system matrix in each iteration. For better efficiency, we solve an equivalent problem instead~\cite{Fang2009}:
\begin{equation}
	(\theta_1^\ast, \ldots, \theta_m^\ast) = \argmin_{ \theta_1, \ldots, \theta_m }~ \left\|F^k - \sum_{j=1}^m \theta_j \Delta F^{k-j}\right\|^2,
	\label{eq:EquivalentLS}
\end{equation}
where $\Delta F^{i} = F^{i+1} - F^i$.  Accordingly, the new iterate becomes
\begin{equation}
\mathbf{Q}_{\mathrm{AA}}^{k+1} = G(\mathbf{Q}^k) - \sum_{j=1}^m \theta_j^\ast \Delta G^{k-j},
\label{eq:AAUpdateFinal}
\end{equation}
where $\Delta G^{k-j} = G(\mathbf{Q}^{k-j+1}) - G(\mathbf{Q}^{k-j})$.
To account for potential linear dependence between $\{\Delta F^{k-j}\}$, we solve the least-squares problem~\eqref{eq:EquivalentLS} by constructing the normal equations
\[
	\left(\mathbf{D}^T \mathbf{D} \right) \boldsymbol{\theta} = \mathbf{D}^T F^k
\]
where $\mathbf{D} = [ \Delta F^{k-1}, \ldots, \Delta F^{k-m} ]$, and computing its minimum-norm solution using complete orthogonal decomposition. 
In this way, the cost of determining $\mathbf{Q}_{\mathrm{AA}}^{k+1}$ amounts to: (i)~computing the latest difference vectors $\Delta F^{k-1}, \Delta G^{k-1}$; (ii)~updating the matrix and the right-hand-side of the normal equation system using $2m$ inner products; and (iii)~solving a small $m \times m$ linear system for $(\theta_1^\ast, \ldots, \theta_m^\ast)$ and applying the result to Equation~\eqref{eq:AAUpdateFinal}. This is typically a small cost compared with the local and global steps.

\subsection{Improving stability}
\begin{figure}[t]
  \centering
  \includegraphics[width=\columnwidth]{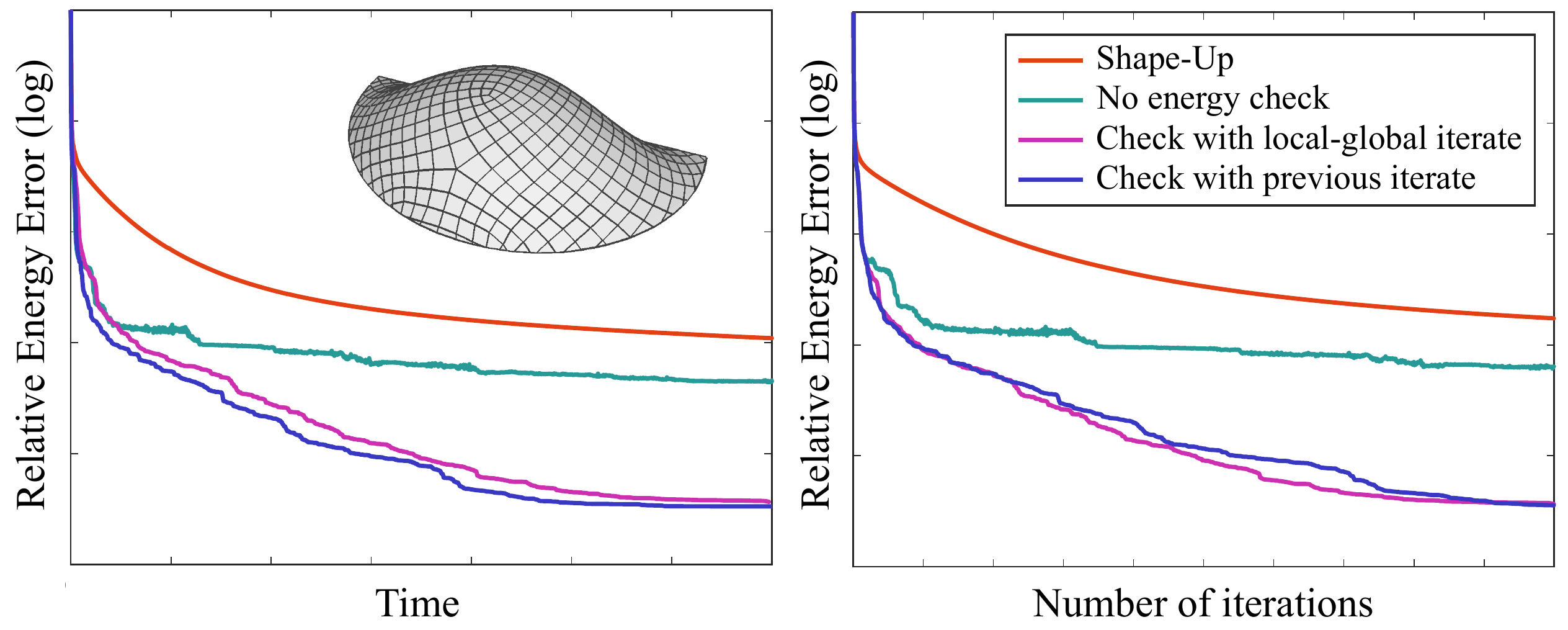}
  \caption{Comparison of target energy decrease between different energy checking strategies for selecting the next iterate, for a PQ mesh optimization problem~\eqref{eq:Planarization}.}
\label{Fig:strategy}
\end{figure}

It has been proved~\cite{Toth2015,Toth2017} that when started from a point close to the solution, Anderson acceleration is convergent under mild conditions. On the other hand, when the iterates are far away from the solution, Anderson acceleration can become unstable, and may lead to slow convergence or stagnation at a wrong solution~\cite{Walker2011,Potra2013}. One example is shown in Figure~\ref{Fig:strategy}, where Anderson acceleration results in oscillation and slow decrease of the target energy when started at a point far from the solution. 

To improve stability, we note that without acceleration, the local-global solver produces an iterate $\mathbf{Q}_{\mathrm{LG}}^{k+1}$ that is guaranteed to decrease the target energy. Therefore, we can compare the target energy values between the accelerated iterate $\mathbf{Q}_{\mathrm{AA}}^{k+1}$ with the unaccelerated one $\mathbf{Q}_{\mathrm{LG}}^{k+1}$, and choose the one with the smaller energy as the next iterate. With this strategy, each iteration decreases the target energy at least as much as the original local-global solver, and the sequence is guaranteed to converge to a local minimum since the target energy is bounded from below. However, it requires two additional evaluations of the target energy at each iteration. Since computing the energy involves projecting the iterate onto all the feasible sets, this can cause a noticeable increase of the computational cost of each iteration. To balance the decrease of total iteration count and the increase of per-iteration cost, we simply compare the target energy of $\mathbf{Q}_{\mathrm{AA}}^{k+1}$ with that of the previous iterate $\mathbf{Q}^k$. If $\mathbf{Q}_{\mathrm{AA}}^{k+1}$ decreases the energy, then it is chosen as the new iterate; otherwise, we choose $\mathbf{Q}_{\mathrm{LG}}^{k+1}$. This strategy only requires one additional energy evaluation if $\mathbf{Q}_{\mathrm{AA}}^{k+1}$ decreases the energy, which is the case for the majority of iterations in our experiments. Moreover, in this case the projections computed for energy evaluation can be reused for performing the global step. Thus the increased computational cost for the selection boils down to evaluating the Euclidean distance between $\mathbf{Q}_{\mathrm{AA}}^{k+1}$ and the projections, which is often negligible. Figure~\ref{Fig:strategy} shows the performance of different selection strategies, using the decrease of target energy with respective to the iteration count and computational time. Within the same computational time, the strategy of comparing with the previous iterate results in the faster decrease of the energy, thanks to its low computational cost per iteration. The full acceleration algorithm with this selection strategy is shown in Algorithm~\ref{algo:AA}.

\begin{algorithm}[t]
	\KwData{\quad $\mathbf{Q}^0$: initial node positions; \\
		\quad $\localstep~(\cdot)$: local step of projection onto feasible sets; \\
		\quad $\globalstep~(\cdot)$: global step of updating node positions;\\
		\quad $G$: the mapping combining the local and global steps; \\
		\quad $E$: the target energy function;\\
		\quad $m$: number of previous iterates used for acceleration.
	}
	\KwResult{A sequence $\{\mathbf{Q}^k\}$ converging to a local minimum of $E$.}
	\BlankLine 
	$\mathbf{Q}^1 = G(\mathbf{Q}^0)$;{~~} $F^0 = \mathbf{Q}^1 - \mathbf{Q}^0$; {~~} $E_{\mathrm{prev}} = +\infty$\;
	\For{$k = 1, 2, \ldots$}{
		\tcp{Make sure $\mathbf{Q}^k$ decreases the energy}
		$\mathbf{P} = \localstep~(\mathbf{Q}^k)$\;
		\If{$E(\mathbf{Q}^k, \mathbf{P}) \geq E_{\mathrm{prev}}$} 
		{
			$\mathbf{Q}^k = \mathbf{Q}_{\mathrm{LG}}$; \quad
			$\mathbf{P} = \localstep~(\mathbf{Q}_{\mathrm{LG}})$;
		}
		$E_{\mathrm{prev}} = E(\mathbf{Q}^k, \mathbf{P})$\;
		\BlankLine 
		\tcp{Anderson acceleration}
		$\mathbf{Q}_{\mathrm{LG}} = \globalstep~(\mathbf{P})$\; 
		$G^k = \mathbf{Q}_{\mathrm{LG}}$; {~~} 
		$F^k = G^k - \mathbf{Q}^{k}$; {~~}
		$m_k = \min(m, k)$\;
		$(\theta_1^\ast, \ldots, \theta_{m_k}^\ast) = \argmin \| F^k - \sum_{j=1}^{m_k} \theta_j \Delta F^{k-j} \|^2$\;
		$\mathbf{Q}^{k+1}  = G^k - \sum_{j=1}^{m_k} \theta_j^\ast \Delta G^{k-j}$;
	}
	\caption{Anderson acceleration for the local-global solver.}
	\label{algo:AA}
\end{algorithm}

\subsection{Choice of $m$}
\label{sec:WindowSizeChoice}
\begin{figure}[t]
	\centering
	\includegraphics[width=\columnwidth]{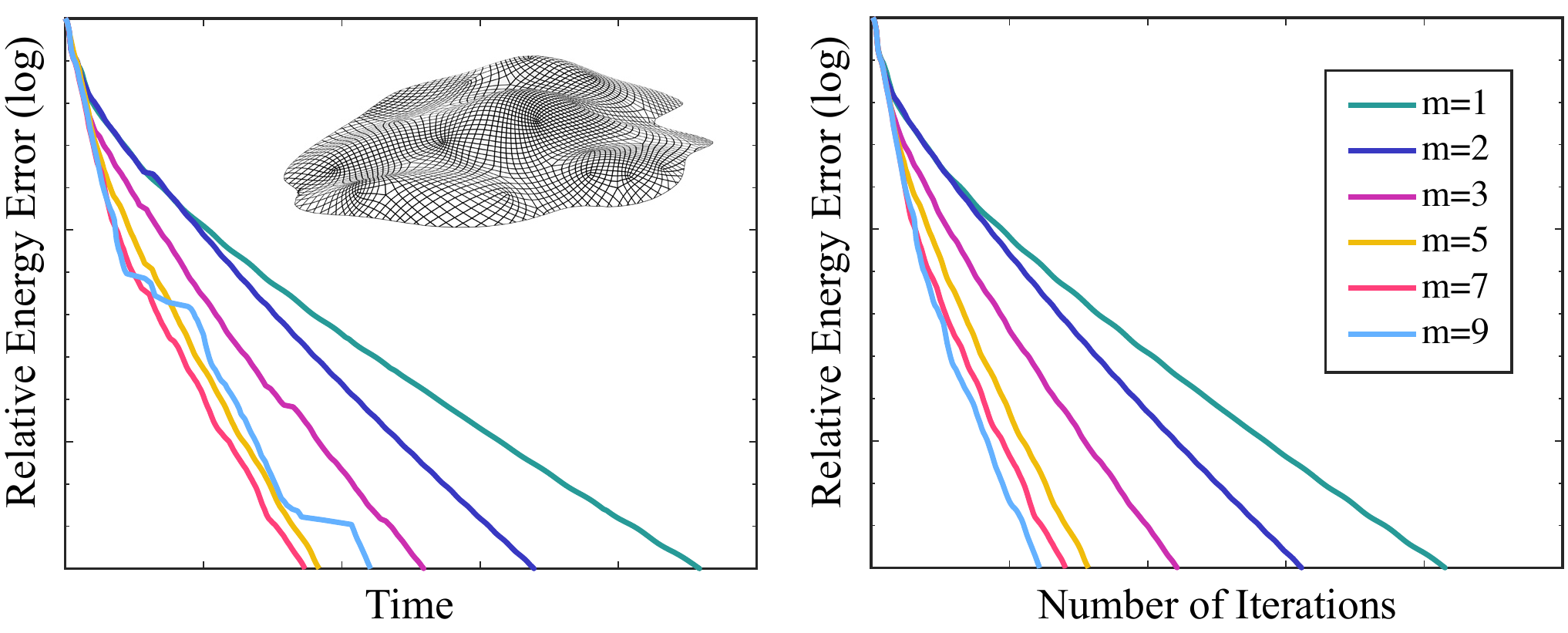}
	\caption{Effect of different values of $m$ for a PQ mesh optimization problem.}
	\label{Fig:ChoiceM}
\end{figure}

The number of previous iterates used for the acceleration (the parameter $m$) has an influence on its performance. With a larger value of $m$, more information is utilized for approximating the inverse Jacobian, which can lead to faster convergence. On the other hand, a larger $m$ increases the computational cost, and may suffer from overfitting iterates that are far away. We observe that beyond $m = 6$, increasing $m$ brings limited improvement in convergence (Figure~\ref{Fig:ChoiceM}). This is consistent with empirical evidence from the literature~\cite{Higham2016}. Therefore, we choose $m = 5$ by default.

\begin{figure*}[t]
	\centering
	\includegraphics[width=\textwidth]{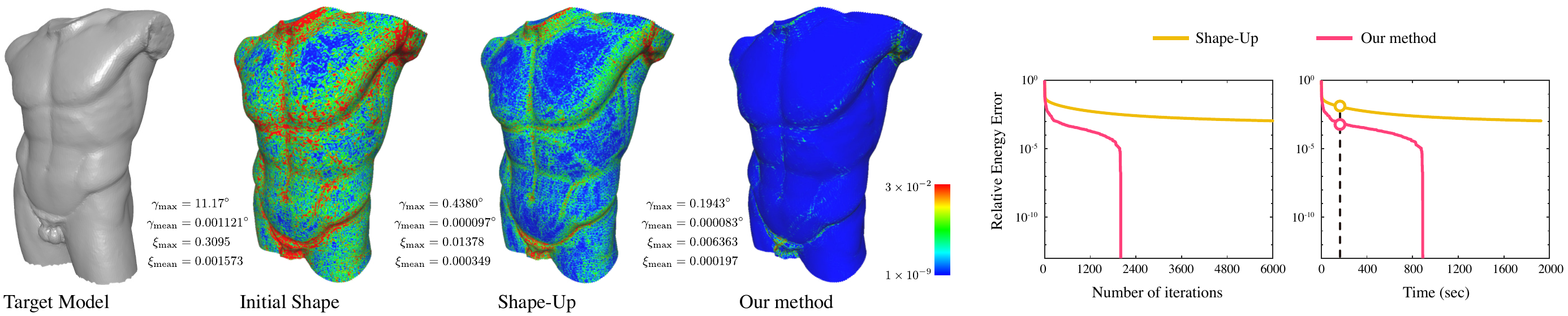}
	\caption{Acceleration of wire mesh optimization~\eqref{eq:WireMesh}, on a model with 230K vertices. The graphs show the relative error of target energy with respect to iterations and computational time. The color-coded models show the distance from each vertex to the reference shape, normalized by the average edge length. The models for Shape-Up and our method are the results after the same computational time, marked using circles on the bottom-right graph. Also shown are the following error metrics: $\xi_{\mathrm{max}}, \xi_{\mathrm{mean}}$: the maximum and average violation of target edge length, normalized using the target edge length; $\gamma_{\mathrm{max}}, \gamma_{\mathrm{mean}}$: the maximum and average violation of angle range constraint in degrees.}
	\label{Fig:wiremesh}
\end{figure*}

\subsection{Analysis}
\label{sec:Analysis}
In all examples so far, we observe significant speed-up from Anderson acceleration which sets it apart from first-order methods. In fact, it has been shown~\cite{Eyert1996,Fang2009,Rohwedder2011} that Anderson acceleration is a quasi-Newton method for solving the nonlinear equation 
\[
	F(\mathbf{Q}) =  {G}(\mathbf{Q}) - \mathbf{Q} = \mathbf{0}.
\]
In particular, computing the accelerated iterate $\mathbf{Q}_{\mathrm{AA}}^{k+1}$ according to Equation~\eqref{eq:FullAAUpdate} is equivalent to~\cite{Fang2009}:
\begin{equation}
	\mathbf{Q}_{\mathrm{AA}}^{k+1} = \mathbf{Q}^{k} + \mathbf{G}_k F(\mathbf{Q}^k), 
	\label{eq:NewtonStep}
\end{equation}
where $\mathbf{G}_k$ is an approximation of the inverse Jacobian of $F$ at $\mathbf{Q}^k$, and is the solution to the following problem
\begin{align}
	\min_{\mathbf{G}}\quad& \|\mathbf{G} + \beta \mathbf{I}\|_F^2 \label{eq:G_Target}\\
	\textrm{s.t.}\quad& {\mathbf{G}} \Delta F^{j} = \Delta {Q}^{j}, \quad j = k-1, \ldots, k-m,
	\label{eq:Secant}
\end{align}
where $\Delta F^{j} = F(\mathbf{Q}^{j+1}) - F(\mathbf{Q}^{j})$, $\Delta \mathbf{Q}^j = \mathbf{Q}^{j+1} - \mathbf{Q}^j$, and $\mathbf{I}$ is the identity matrix. Note that the inverse Jacobian should map the differential of $F$ to the differential of $\mathbf{Q}$. This condition is enforced using the \emph{secant equations}~\eqref{eq:Secant}, where the differentials are approximated using finite difference between previous iterates. The target function~\eqref{eq:G_Target} measures the difference between the inverse Jacobian and the matrix $-\beta \mathbf{I}$. Intuitively, this means we obtain $\mathbf{G}^k$ by taking $-\beta \mathbf{I}$ as an initial approximation of the inverse Jacobian, and applying minimum modification to make it satisfy the secant equations and better capture the variation $F$ around the current iterate. $\mathbf{G}^k$ is then utilized in a Newton step~\eqref{eq:NewtonStep} to bring the function $F$ closer to zero.
 
This perspective not only explains the fast convergence we observe in the experiments, but also justifies our choice of mixing parameter $\beta = 1$. In this case, the inverse Jacobian is initially approximated by $- \mathbf{I}$, which implies vanishing Jacobian of the fixed-point mapping $G$ at $\mathbf{Q}^k$. This condition is met if the mapping between node positions $\mathbf{Q}$ and projection variables $\mathbf{P}$ has vanishing Jacobian, e.g. if we fix $\mathbf{P}$ at the projection of $\mathbf{Q}^k$. This is the same assumption made by the local-global solver to reduce the global step into a simple linear solve. Since the Jacobian of projection points indicates the curvature of the feasible sets, our choice of $\beta = 1$ means we start from an inverse Jacobian approximation that has no prior knowledge of the feasible sets, and then use the previous iterates to infer and add in curvature information.

Indeed, the same strategy has been used in~\cite{LiuBK17} to construct an initial descent direction for their L-BFGS solver of projective dynamics. The goal of their solver is to find a root of the equation $\mathbf{g}(\mathbf{Q}) = \mathbf{0}$, where $\mathbf{g}$ is the gradient of the projective dynamics energy~\eqref{eq:ProjDyna}. Each iteration computes a descent direction and performs line search to determine the next iterate. To compute the descent direction, they start with an initial estimate $-\mathbf{H} \mathbf{g}(\mathbf{Q}^k)$ where $\mathbf{H}$ is an approximate inverse Hessian of the target function, followed by a two-loop recursion that implicitly updates the inverse Hessian and the descent direction according to $m$ previous iterates. In~\cite{LiuBK17}, the initial inverse Hessian $\mathbf{H}$ is chosen to be the inverse of matrix $\mathbf{M}/h^2 + \sum_i \mathbf{A}_i^T \mathbf{A}_i$ of the global step~\eqref{eq:Physics}, which is the Hessian of a modified target energy with the projection variables $\mathbf{P}$ fixed. In other words, their initial inverse Hessian is constructed using the same assumption as for our initial inverse Jacobian. From this point of view, the main difference between our technique and~\cite{LiuBK17} is in the update of the inverse Hessian/Jacobian. Anderson acceleration modifies the inverse Jacobian by enforcing the $m$ secant equations simultaneously, while the L-BFGS two-loop recursion updates the inverse Hessian in $m$ steps, each applying minimum modification to satisfy one secant equation~\cite{nocedal2006numerical}. In general, the L-BFGS inverse Hessian is not guaranteed to satisfy all $m$ secant equations except for the last one. In addition, this leads to different computational costs between the two methods: for both of them, the cost is dominated by inner products between vectors of the same length; the two-step recursion requires $2m+1$ inner products to be done sequentially, while our method requires only $m$ inner products that can be parallelized. As a result, our method incurs a lower computational cost than L-BFGS.

From another perspective, our selection strategy for stabilizing Anderson acceleration has a similar effect as the line search employed by~\cite{LiuBK17} to ensure stability of L-BFGS. In~\cite{LiuBK17}, the quasi-Newton step is guaranteed to be along a descent direction, but the default step size may actually increase the target energy. Thus they adjust the step size using backtracking line search, to guarantee sufficient decrease of energy by satisfying the Armijo condition. Similarly, when an Anderson acceleration iterate increases the energy value, we revert to the local-global iterate; this can be seen as a single-step line search that guarantees energy decrease. Unlike the line search in~\cite{LiuBK17}, our strategy does not guarantee the Armijo condition, which in theory may not reduce the energy as much as L-BFGS in some cases. On the other hand, our single-step line search guarantees low computational overhead, which allows for more iterations within the same time. In practice, we observe similar decrease of energy per iteration between our method and L-BFGS (see Section~\ref{sec:Results}).

\subsection{Beyond local-global solvers}
Our method is applicable to other iterative solvers, as long as they monotonically decrease the target energy, and there is a well-defined mapping between two consecutive iterates. In some cases, even solvers that do not strictly satisfy these conditions can be accelerated with minor modification. Some examples are provided in Section~\ref{sec:OtherSolvers}.

\begin{figure*}[t]
	\centering
	\includegraphics[width=\textwidth]{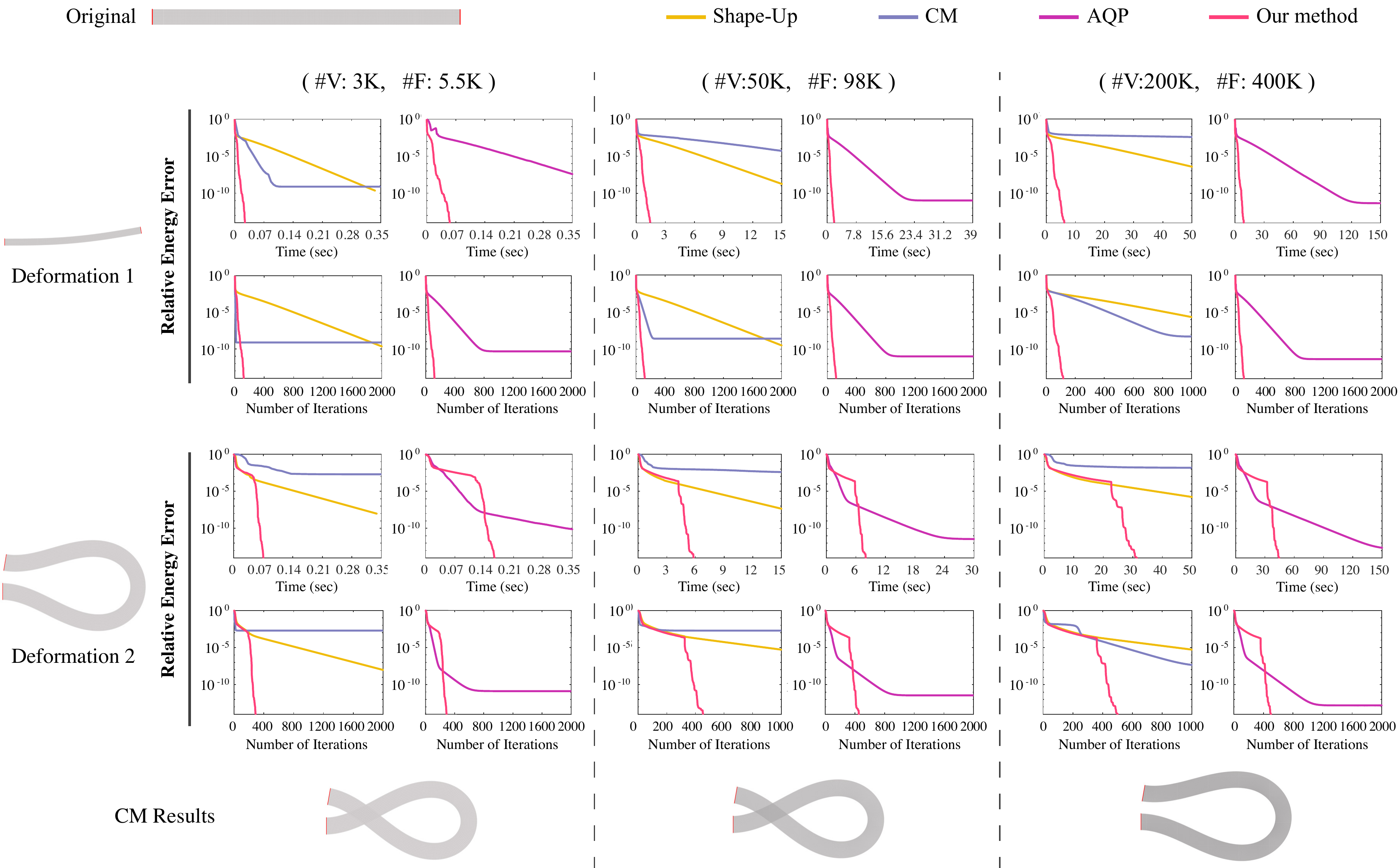}
	\caption{Comparison between different solvers for 2D ARAP deformation of a bar model in different resolutions, by plotting their relative energy errors. The vertices at two ends of the bar are used as handles (shown in red). For fair comparison, we compare AQP with a single-threaded implementation of our algorithm. Note that the CM solver may converge to a suboptimal local minimum, as shown in the last row.}
	\label{Fig:2DARAP}
\end{figure*}
\section{Results}
\label{sec:Results}

In this section, we evaluate the behavior and performance of our acceleration technique on a variety of geometry optimization and physics simulation problems, and compare it with existing methods. We solve the geometry optimization problems using an accelerated version of the Shape-Up solver~\cite{BouazizDSWP12}, and the physics simulation problems using an accelerated Projective Dynamics solver~\cite{Bouaziz2014}.
We compare different methods by starting from the same initial solution, and plotting for each method the graphs of relative energy error $\overline{E}$ or the relative distance error $\overline{D}$ with respect to the computational time and iteration counts. The relative energy error is defined as 
\begin{equation}
	\overline{E} = \frac{E - E^\ast}{E^0 - E^\ast}.
	\label{eq:RelEnergy}
\end{equation}	
Here $E$ is the current energy value, and $E^0$ is the energy value for the initial solution; $E^\ast$ is the energy value at the final solution, computed by running all methods until full convergence and taking the lowest energy value among their final results. Similarly, the relative distance error is defined as 
\begin{equation}
	\overline{D} = \frac{\|\mathbf{Q} - \mathbf{Q}^\ast\|}{\|\mathbf{Q}^0 - \mathbf{Q}^\ast\|},
\end{equation}	
where $\mathbf{Q}, \mathbf{Q}^0$ are the current and initial solutions, and $\mathbf{Q}^\ast$ is the final solution corresponding to the final energy $E^\ast$ in Eq.~\eqref{eq:RelEnergy}.

Our algorithm is implemented in C++, using \eigen{}~\cite{eigenweb} for linear algebra, and \libigl~\cite{libigl} for geometry processing operations. Unless stated otherwise, all examples are run on a desktop PC with 16GB of RAM and a quad-core CPU at 3.6 GHz. For the best performance, all methods are parallelized using OpenMP where appropriate.
The source code of our method is available at~{\url{https://github.com/bldeng/AASolver}}.

\subsection{Geometry Optimization}

\para{Planarization} Figure~\ref{Fig:Planarity} shows an example of planar quadrilateral (PQ) mesh optimization, which is an important problem for freeform architectural design~\cite{Liu2006}. Starting from the initial quad mesh and a reference triangle mesh, we planarize the quad mesh by minimizing a target energy about its vertex positions:
\begin{equation}
	E_1 = w_{\mathrm{planar}} E_{\mathrm{planar}} + w_{\mathrm{ref}} E_{\mathrm{ref}} + w_{\mathrm{fair}} E_{\mathrm{fair}} + w_{\mathrm{2nd}} E_{\mathrm{2nd}},
	\label{eq:Planarization}
\end{equation}
where $w_{\mathrm{planar}}, w_{\mathrm{ref}}, w_{\mathrm{fair}}, w_{\mathrm{2nd}}$ are positive weights, $E_{\mathrm{planar}}$ is a planarity term that measures the sum of squared distance between each face and its best fitting plane~\cite{BouazizDSWP12}, $E_{\mathrm{ref}}$ is a reference term that sums the squared distance from each vertex to the reference mesh, and $E_{\mathrm{fair}}, E_{\mathrm{2nd}}$ are Laplacian fairness and relative Laplacian fairness terms as defined in~\cite{Liu2011}. The projection operators for the planarity terms are computed via SVD~\cite{BouazizDSWP12}, while the projection operators of the reference term are evaluated by finding the closest point on the reference surface from each quad mesh vertex, and implemented using an AABB tree. Figure~\ref{Fig:Planarity} shows the relative energy graphs for the Shape-Up solver~\cite{BouazizDSWP12} and its accelerated version using our technique. 
The Shape-Up solver suffers from slow convergence after the initial iterations, while the accelerated solver requires significantly fewer iterations and less computational time to achieve the final solution.

\begin{figure*}[t!]
	\centering 
	\includegraphics[width=\textwidth]{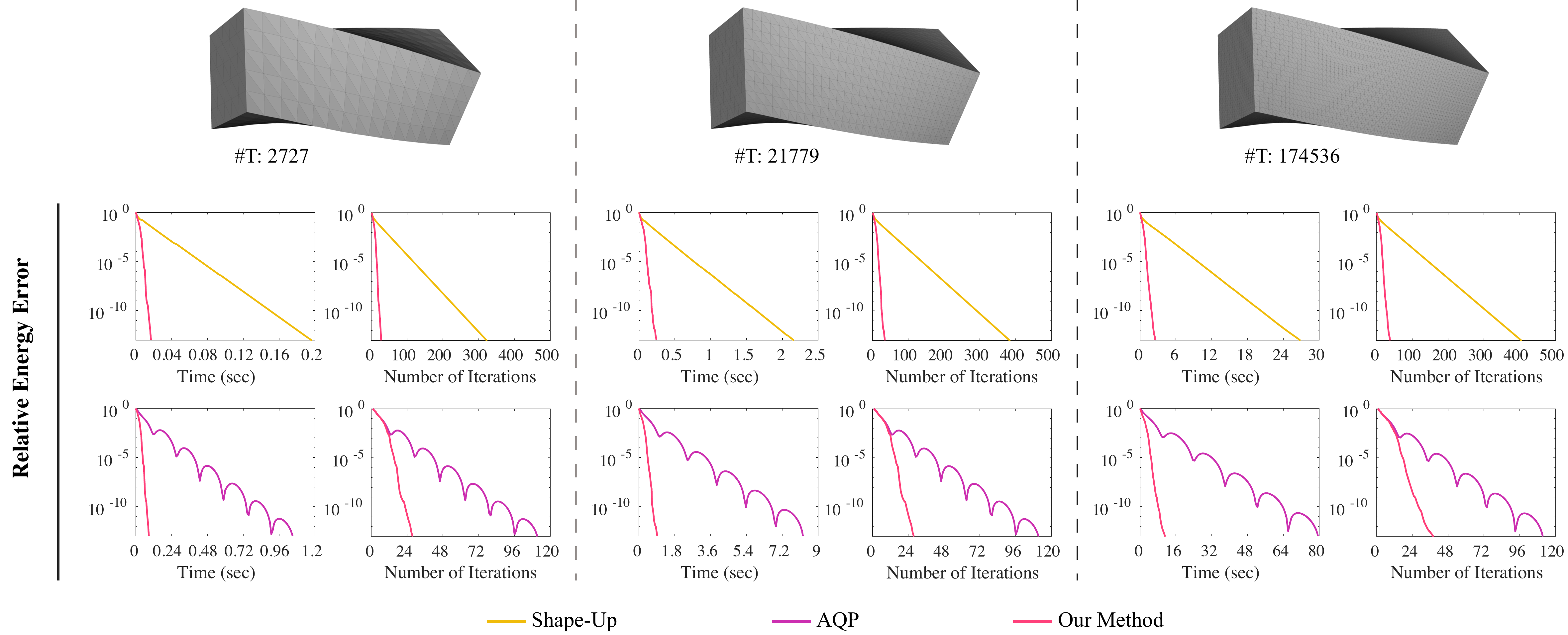}
	\caption{ARAP deformation of a 3D bar in three different resolutions, using vertices at the two ends as handles.}
	\label{Fig:3DBar}
\end{figure*}

\para{Wire mesh design} Figure~\ref{Fig:wiremesh} shows an example of material-aware design, where a wire mesh model is optimized to approximate a target surface~\cite{Grinspun14b}. The wire mesh is represented as a quadrilateral mesh, subject to the following geometric constraints that model its material properties: (i) all edge lengths equal to a constant $l$; (2) within each face, all four corner angles are between 45 and 135 degrees~\cite{Grinspun14b}. Given a reference surface and an initial wire mesh shape, we compute the final wire mesh shape by minimizing the target function
\begin{equation}
	E_2 = w_{\mathrm{edge}}  E_{\mathrm{edge}}  + w_{\mathrm{angle}}  E_{\mathrm{angle}} + w_{\mathrm{ref}} E_{\mathrm{ref}},
	\label{eq:WireMesh}
\end{equation}
where $E_{\mathrm{ref}}$ is a reference surface term defined in the same way as in Equation~\eqref{eq:Planarization}. $E_{\mathrm{edge}}, E_{\mathrm{angle}}$ are shape proximity terms for the edge length constraints and the angle constraints, respectively:
\begin{align}
	E_{\mathrm{edge}} &= \sum_{i \in \mathcal{E}} \| \mathbf{q}_{i_1} - \mathbf{q}_{i_2} - \mathbf{p}_i\|^2 + \sigma_{\mathrm{edge}}(\mathbf{p}_i),\label{eq:EdgeLengthTerm}\\
	E_{\mathrm{angle}} &= \sum_{(i,j,k) \in \mathcal{A}} \left\| 
	\begin{bmatrix} 
		\mathbf{q}_{i} - \mathbf{q}_j \\
		\mathbf{q}_{k} - \mathbf{q}_j
	\end{bmatrix}
	-
	\begin{bmatrix} 
		\mathbf{e}_{ijk}^1 \\
		\mathbf{e}_{ijk}^2
	\end{bmatrix}
	\right\|_F^2
	+ \sigma_{\mathrm{angle}}
	\left(	 
	\begin{bmatrix} 
	\mathbf{e}_{ijk}^1 \\
	\mathbf{e}_{ijk}^2
	\end{bmatrix}
	\right),
\end{align} 
where $\mathcal{E}$ is the index set of mesh edges, $\mathbf{q}_{i_1}$ and $\mathbf{q}_{i_2}$ are the vertices of edge $i$, $\sigma_{\mathrm{edge}}$ is the indicator function for the edge length constraint feasible set $\{\mathbf{e}\mid\|\mathbf{e}\| = l\}$; $\mathcal{A}$ is the index set of vertices that form a face corner, and $\sigma_{\mathrm{angle}}$ is the indicator function for the angle constraint feasible set $\left\{ (\mathbf{e}_1, \mathbf{e}_2) \left|~ \cos(\displaystyle\frac{3\pi}{4}) \leq \displaystyle\frac{\mathbf{e}_1}{\|\mathbf{e}_1\|} \cdot \displaystyle\frac{\mathbf{e}_2}{\|\mathbf{e}_2\|} \leq \cos(\displaystyle\frac{\pi}{4}) \right.\right\}$. The projection operators for the edge length constraint and the angle constraint are given in~\cite{Deng2015}.
The accelerated solver only takes a fraction of iterations to compute an accurate result compared with the unaccelerated one, which significantly reduces the computational time.

\begin{figure}[b!]
	\centering
	\includegraphics[width=\columnwidth]{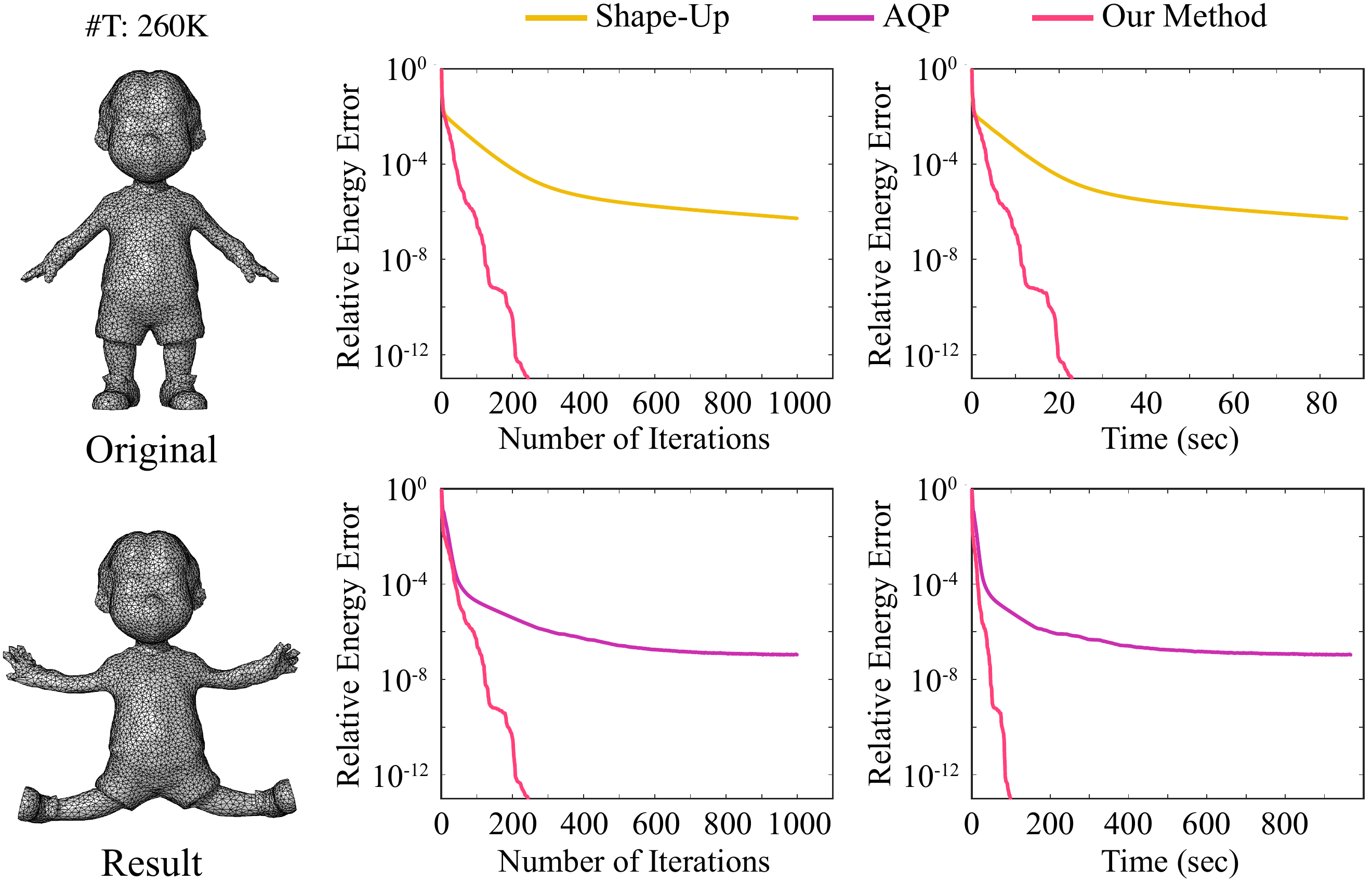}
	\caption{3D ARAP deformation of a mesh, with handles located at the limbs.}
	\label{Fig:3DARAP}
\end{figure}

\begin{figure*}[t]
	\centering
	\includegraphics[width=\textwidth]{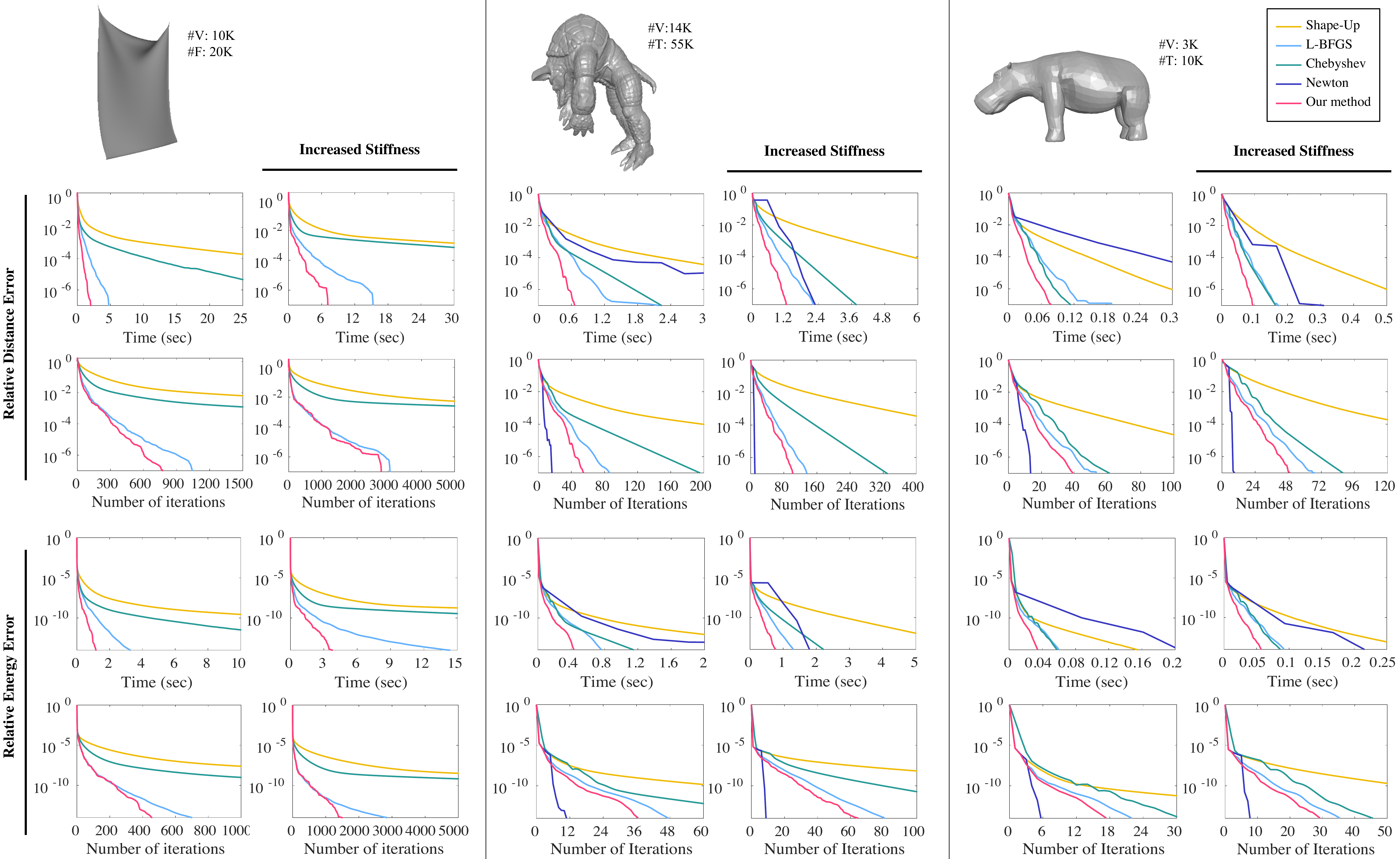}
	\caption{Acceleration of physics simulation. The graphs show in log scale the relative energy error (top two rows) and the relative distance error (bottom two rows) for each solver. Each model is tested with two stiffness settings, and the results with increased stiffness are shown on the right column.}
	\label{Fig:SimulationGraph}
\end{figure*}

\begin{figure}[t!]
	\centering
	\includegraphics[width=1\columnwidth]{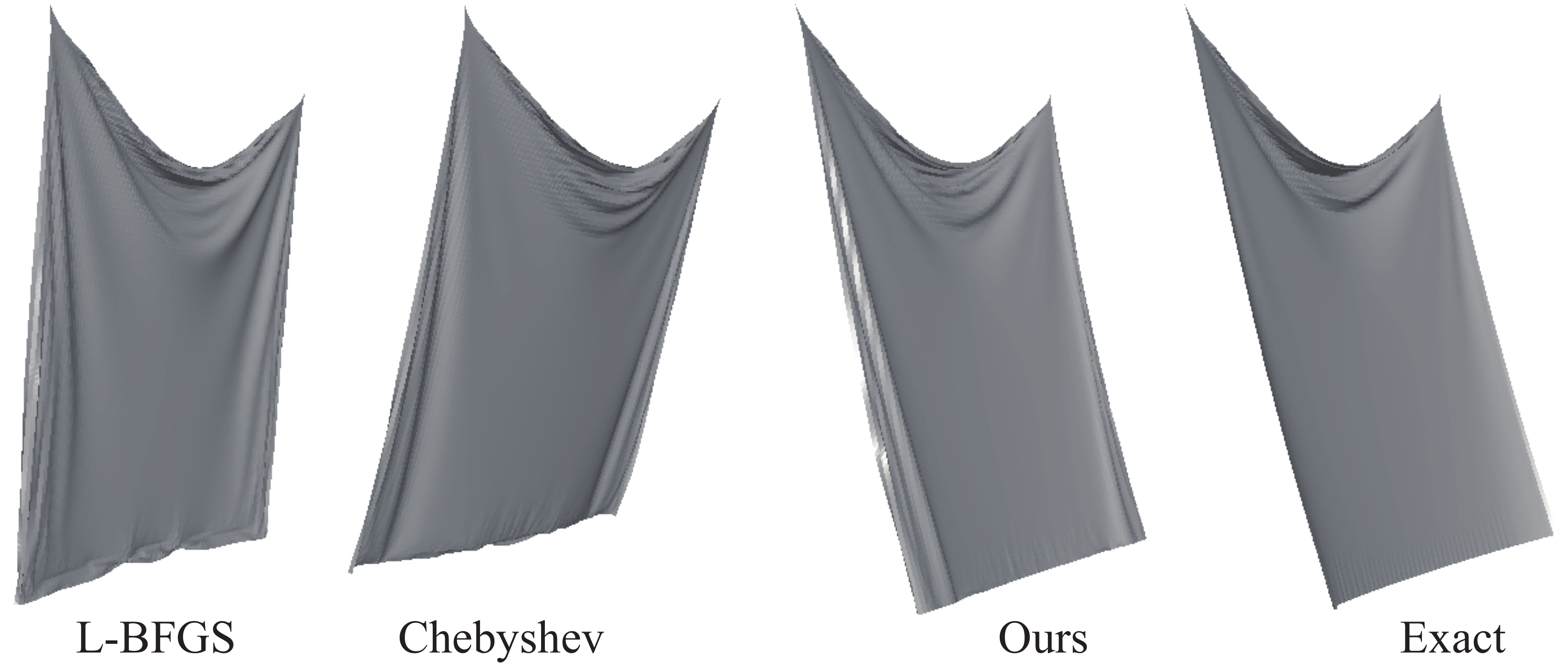}
	\caption{Comparison of a cloth simulation frame at the same time instance, computed using different methods with the same computational time budget per frame. The result using our method is closer to the exact solution of the simulation sequence.}
	\label{fig:SimulationComparison}
\end{figure}

\para{ARAP deformation} In Figures~\ref{Fig:2DARAP} and~\ref{Fig:3DARAP}, we perform ARAP deformation of 2D triangle meshes and 3D tetrahedron meshes according to user-driven handle vertices, by solving the problem subject to hard constraints of handle vertex positions:
\begin{equation}
	\min_{\mathbf{Q}_{\textrm{free}}} \sum_{i \in \mathcal{F}} \left\| ~\mathbf{J}_i \mathbf{Q}_i - \mathbf{R}_i \right\|_F^2 + \sigma_{\mathrm{rot}} (\mathbf{R}_i),
	\label{eq:ARAPEnergy}
\end{equation}
where $\mathbf{Q}_{\textrm{free}}$ are the positions of non-handle vertices, $\mathcal{F}$ is the index set of triangles or tetrahedrons, $\mathbf{Q}_i \in \mathbb{R}^{{d+1} \times d }$ collects the positions of the $d+1$ vertices in cell $i$, $\mathbf{J}_i \in \mathbb{R}^{d \times (d+1)}$ is a linear map that computes the deformation gradient of cell $i$ between positions $\mathbf{Q}_i$ and initial positions ${\mathbf{Q}}_i^0$, and $\sigma_{\mathrm{rot}}$ is the indicator function for $d \times d$ rotation matrices. With the Shape-Up solver, the projection operator in the local step finds the closest rotation matrix to a given matrix, which we compute using SVD according to~\cite{SorkineRabinovich2016}. To enforce handle positions as hard constraints, we modify the global step of the Shape-Up solver by removing the rows that correspond to handle vertices. 
We compare the performance of our accelerated solver with other ARAP solvers that support hard-constraint handles, including the accelerated quadratic proxy (AQP) solver from~\cite{KovalskyGL16}, and the composite majorization (CM) solver from~\cite{ShtengelPSKL17}. 
The CM formulation of 2D ARAP deformation is derived by representing the ARAP energy as a function of singular values for the Jacobian, and using the singular value formulas provided in~\cite{ShtengelPSKL17}.
The AQP solver and the CM solver are implemented using the source codes provided by the authors\footnote{\url{https://github.com/shaharkov/AcceleratedQuadraticProxy}}\footnote{\url{https://github.com/Roipo/CompMajor}}. Since the AQP code is implemented using a single thread, we compare with it with a single-threaded implementation of our solver for fair comparison.
Figure~\ref{Fig:2DARAP} shows the deformation of a 2D bar model represented using meshes of different resolutions and with different target handle positions. 
Figure~\ref{Fig:3DBar} shows the deformation of a 3D bar model with different resolutions.
Figure~\ref{Fig:3DARAP} shows the deformation of a 3D mesh with 260K tetrahedrons. 
We can see that depending on the model and the configuration, different solvers may converge to different local minima. In most cases, our accelerated solver converges to the lowest energy with small computational cost.

\subsection{Physics simulation}

\begin{figure*}[t]
	\centering
	\includegraphics[width=\textwidth]{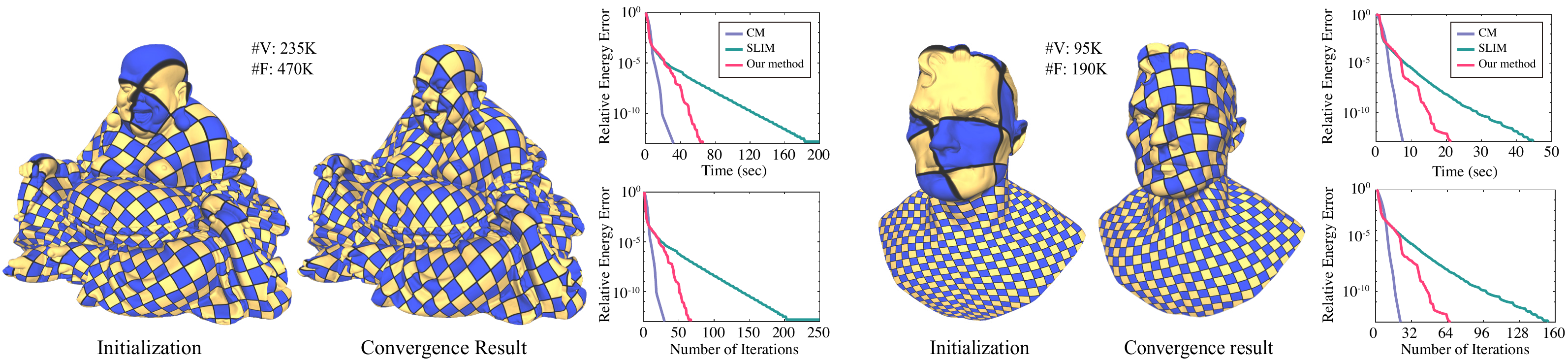}
	\caption{Optimizing the symmetric Dirichlet energy, using the original SLIM algorithm, our accelerated version, and the CM solver.}
	\label{fig:slim}
\end{figure*}

\begin{figure}[b!]
	\centering
	\includegraphics[width=1\columnwidth]{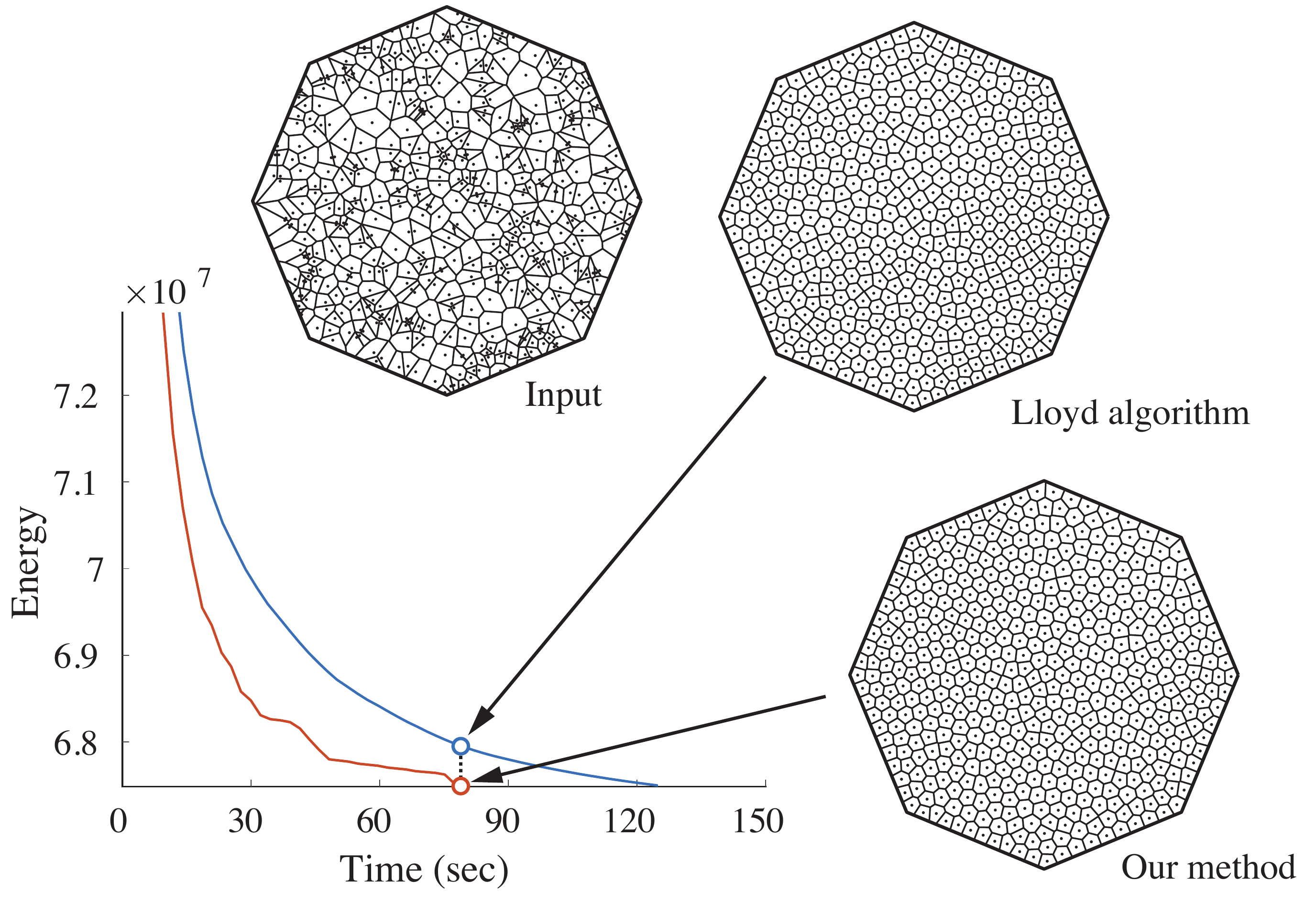}
	\caption{Acceleration of the Lloyd algorithm for minimizing the CVT energy on a Octagon shape boundary and 500 generating points.}
	\label{fig:cvt}
\end{figure}

We apply our accelerated Projective Dynamics solver to simulate the deformation of cloth and elastic solids under gravity and subject to positional constraints. We model a piece of cloth as a mass-spring network represented as a triangle mesh, where each edge is subject to a length constraint with its rest length being the target length. Such constraint defines an elastic potential energy term that has the same form as in Equation~\eqref{eq:EdgeLengthTerm}, weighted by the spring stiffness constant~\cite{Liu2013}. For elastic solids, we represent them as tetrahedron meshes, and define its elastic potential energy in the same way as the target function in~\eqref{eq:ARAPEnergy}, weighted by an elasticity parameter. For each node $\mathbf{q}_i$ that needs to be fixed during simulation, we add a penalty term $0.5 \cdot w_{\mathrm{fixed}} \|\mathbf{q}_i - \overline{\mathbf{q}}_i\|^2$ to the Projective Dynamics target energy~\eqref{eq:ProjDyna}, where $\overline{\mathbf{q}}_i$ is the constrained position and $w_{\mathrm{fixed}}$ is a large positive weight. To evaluate the effectiveness of our acceleration, we compare it with the Chebyshev semi-iterative method in~\cite{Wang15}, and the L-BFGS approach in~\cite{LiuBK17}. We apply the Chebyshev semi-iterative method with a direct solver for the full step instead of the Jacobi solver proposed in the paper, because the direct solver is more efficient on the CPU~\cite{Wang15}. For the L-BFGS solver, the two-loop recursion is performed using five previous iterates as suggested by~\cite{LiuBK17}. 
For elastic solids, we also compare with the Newton solver derived in~\cite{Sifakis2012}. Since the local-global solver converges quickly to an approximate solution and the Newton solver enables local quadratic convergence and the accurate solution, we perform five iterations of local-global solve before switching to the Newton solver, to achieve good performance. The Newton linear system is solved using PARDISO, with symbolic pre-factorization to maximize its efficiency. Each model is tested under two stiffness settings, by setting the spring stiffness constant or the elasticity parameter.
Figure~\ref{Fig:SimulationGraph} shows the decrease of relative energy error and relative distance error for each solver, for the computation of the first frame in the simulation. Due to the very large initial energy, the relative distance error provides better indication of convergence to solution. We can see that all three accelerated solvers perform better than the original projective dynamics solver. Our solver has similar performance with the L-BFGS in the decrease of energy with respect to iteration counts, potentially because the two solvers construct the initial inverse Hessian/Jacobian in a similar way, and both use five previous iterates for acceleration. Our solver performs better than L-BFGS in terms of computational time, due to its lower cost of computing the accelerated iterate. Both solvers perform better than the Chebyshev solver, potentially because the Chebyshev approach is equivalent to quasi-Newton using two previous iterates~\cite{LiuBK17}, thus with less accurate approximation of the Hessian. Although the Newton solver requires the least iterations to converge in most cases, it is not the best-performing solver in terms of computational time, due to its high computational cost per iteration.

Figure~\ref{fig:SimulationComparison} shows a comparison of the simulation results using different solvers with the same computational budget for each frame. The results are compared with an exact solution sequence, where each frame is computed by running the L-BFGS solver until full convergence. The simulation result using our solver is very close the exact solution, while other solvers lead to noticeable difference compared with the exact result. The full simulation results are shown in the accompanying video\footnote{\url{https://www.youtube.com/watch?v=O6mJBlG9-jk}}.

\subsection{Other solvers}
\label{sec:OtherSolvers}

To verify the effectiveness of our acceleration technique beyond the classical local-global solvers, we apply it to two other iterative solvers. The first one is the Lloyd algorithm for computing centroidal Voronoi tessellation (CVT) for a bounded domain, where the generating points of the Voronoi cells coincide with the centroids of the cells~\cite{Du2006}. Given a density function $\rho$ defined on the domain, the generating points $\mathbf{X} = \{\mathbf{x}_i\}$ of a CVT correspond to a minimum of the energy~\cite{Du1999}:
\[
	F(\mathbf{X}) = \sum_{i} \int_{\Omega_i} \rho(\mathbf{y}) \|\mathbf{y} - \mathbf{x}_i\|^2 d \mathbf{y},
\]
where $\Omega_i$ is the Voronoi cell for $\mathbf{x}_i$. The Lloyd algorithm is a fixed-point iteration to minimize this energy
\begin{equation}
	\mathbf{x}_i^{k+1} = \frac{\int_{\Omega_i^k} \mathbf{y} \rho(\mathbf{y}) d\mathbf{y}}{\int_{\Omega_i^k}\rho(\mathbf{y}) d\mathbf{y}},
	\label{eq:LloydCVT}
\end{equation}
where $\Omega_i^k$ is the Voronoi cell for the current iterate $\mathbf{x}_i^k$; i.e, in each iteration the generating points are moved to the centroid of their current Voronoi cells. The Lloyd algorithm is shown to decrease the target energy and converges to a CVT~\cite{Du2006}. However, it suffers from slow convergence to the solution, which has prompted the development of faster CVT solvers based on L-BFGS which require evaluation of the energy gradient~\cite{Liu2009}. Our acceleration technique can be directly applied to the fixed-point iteration~\eqref{eq:LloydCVT} to produce an accelerated iterate $\mathbf{X}_{\mathrm{AA}}^{k+1}$. We accept $\mathbf{X}_{\mathrm{AA}}^{k+1}$ as the new iterate if it decreases the target energy compared with the previous iterate and the generating points are inside the domain boundary; otherwise we revert to the Lloyd algorithm's result as the next iterate. In Figure~\ref{fig:cvt}, we apply the acceleration to a publicly available implementation of Lloyd algorithm\footnote{\url{https://github.com/Narusaki/CVT2D}}. Without evaluating the gradient or using preconditioner, our simple acceleration achieves comparable speed-up to the L-BFGS solver proposed in~\cite{Liu2009}. This example shows a benefit of our acceleration technique: it only requires the evaluation of target energy, and can be easily added on top of existing codes.

\begin{figure}[t!]
	\centering
	\includegraphics[width=\columnwidth]{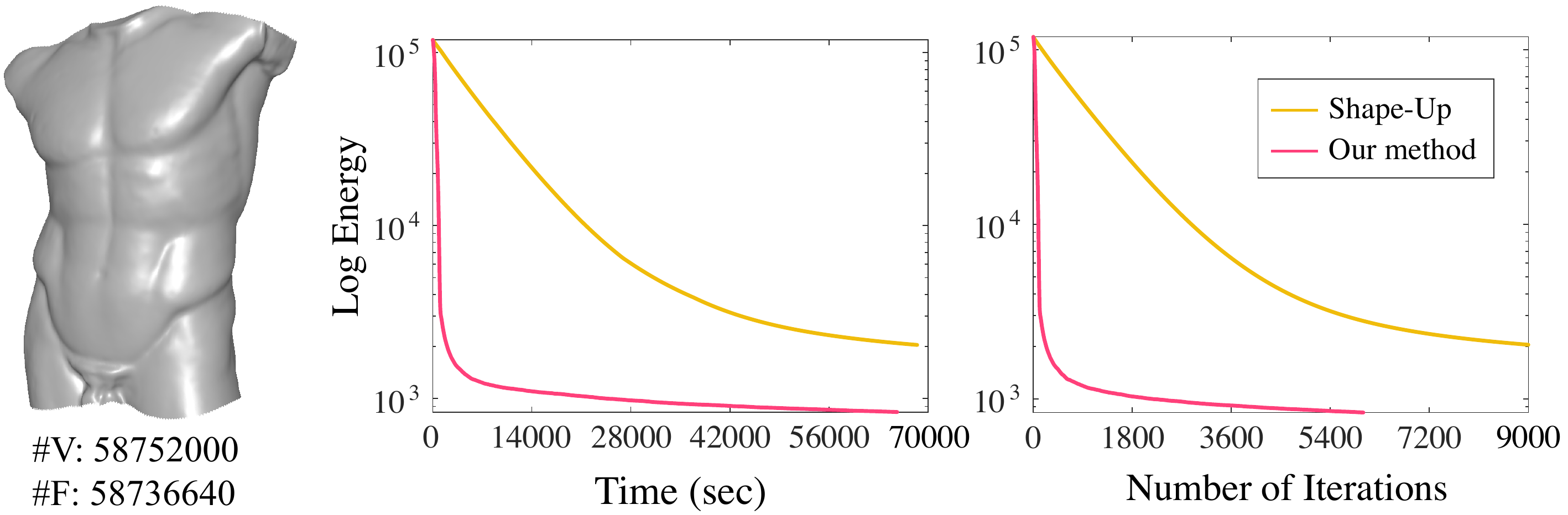}
	\caption{Acceleration of wire mesh optimization on a very large model.}
	\label{fig:StressTestWireMesh}
\end{figure}

Another example is the scalable locally injective mapping (SLIM) solver~\cite{RabinovichPPS17}, which computes injective mesh parameterization by minimizing a flip-preventing energy:
\[
	\min_{\mathbf{x}} \sum_{f \in F} A_f D(~\mathbf{J}_f(\mathbf{x})),
\] 
where $\mathbf{x}$ is the parameterization coordinates at the vertices, $F$ is the mesh face set, $A_f$ is the cell area/volume, $\mathbf{J}_f$ is the parameterization Jacobian on face $f$, and $D$ is a distortion measure. As the energy cannot be directly minimized using classical local-global solvers, the authors adopt a reweighting strategy: the local step remains the same as other local-global solvers, while the global step constructs a reweighted proxy function based on the current iterate $\mathbf{x}^k$ and minimizes it to obtain a solution $\mathbf{p}^{k+1}$ that provides a decent direction. Then a bisection line search is performed to determine the next iterate $\mathbf{x}^{k+1} = \mathbf{x}^k + \alpha (\mathbf{p}^{k+1} - \mathbf{x}^k)$ which reduces the original energy while ensuring injectivity of the mapping, with an initial step size $\alpha = \min\{0.8 \alpha_{\max}, 1\}$ that prevents element flips where $\alpha_{\max}$ is the maximal step size with no foldovers. As mentioned in~\cite{RabinovichPPS17}, this method has similar convergence property to classical local-global solvers, with slow convergence to high-accuracy solutions.
Since the new iterate is determined via discrete line search, there is no well-defined mapping between consecutive iterates $\mathbf{x}^k$ and $\mathbf{x}^{k+1}$, and we cannot directly apply Anderson acceleration to the sequence $\{\mathbf{x}^k\}$. On the other hand, the proxy solution $\mathbf{p}^{k+1}$ is determined by minimizing a convex energy that depends on $\mathbf{x}^{k}$, which can be written as a mapping $\mathbf{p}^{k+1} = G(\mathbf{x}^k)$; moreover, a fixed-point of the mapping $G$ is a local minimum of the original energy. Therefore, we apply Anderson acceleration to the mapping $G$, to determine an accelerated proxy solution $\mathbf{p}_{\mathrm{AA}}^{k+1}$. To choose the next iterate, we evaluate the original target energy using the initial line search steps according to $\mathbf{p}_{\mathrm{AA}}^{k+1}$ and $\mathbf{p}^{k+1}$ respectively, and choose the one with lower energy to continue with the line search. In this way, the acceleration helps to find a descent direction with faster decrease of the original energy. Figure~\ref{fig:slim} compares the performance of the original SLIM algorithm, our accelerated version, and the CM solver, in  optimizing the symmetric Dirichlet energy for parameterization. Our technique provides effective acceleration for SLIM, although still slower than CM.

\subsection{Stress Tests}
We also perform a series of stress tests to evaluate the performance of our method in extreme cases. In Figure~\ref{fig:StressTestWireMesh}, we perform wire mesh optimization with the same settings as in Figure~\ref{Fig:wiremesh}, but on a model with over 50M vertices. The model is derived from the mesh in Figure~\ref{Fig:wiremesh} by repeatedly subdividing each quad face into four quads. For such a large model, it is impractical to fully solve the linear system in the global step. Instead we adopt the strategy from~\cite{Wang15} and only perform one step of Jacobi iteration for the linear system. The experiment was run on a workstation with 48 Xeon cores and 128GB RAM. Since it takes too long for the solvers to achieve full convergence, we only plot the energy graphs instead of the relative energy errors. We can see that our method achieves similar speedup as in the smaller model from Figure~\ref{Fig:wiremesh}.

\begin{figure}[t!]
 	\centering
 	\includegraphics[width=\columnwidth]{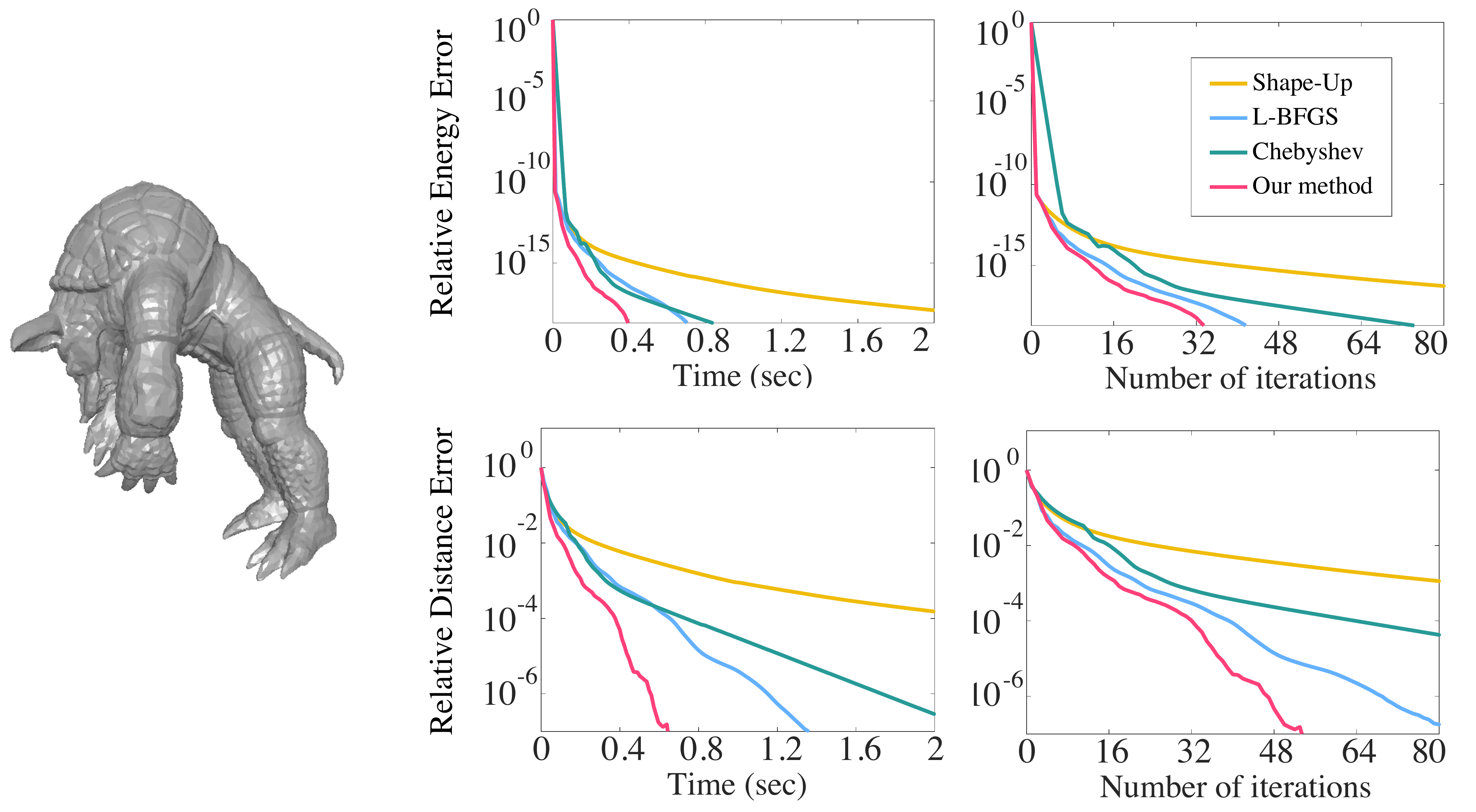}
 	\caption{Physics simulation of an elastic solid with degenerate tetrahedrons. The model is derived by randomly adding 100 degenerate tetrahedrons to the Armadillo model in Figure~\ref{Fig:SimulationGraph}.}
 	\label{fig:DegenerateMesh}
\end{figure}

In Figure~\ref{fig:DegenerateMesh}, we apply our method for physics simulation to a mesh model with degenerate elements. We randomly add 100 degenerate tetrahedrons to the Armadillo model used in Figure~\ref{fig:SimulationComparison}. Each degenerate tetrahedron is added by randomly picking a triangle face, introducing a new vertex on a random position within the face, and connecting the new vertex to the three triangle vertices. Our method remains effective in the presence of degenerate elements.

\begin{figure}[t]
	\centering
	\includegraphics[width=\columnwidth]{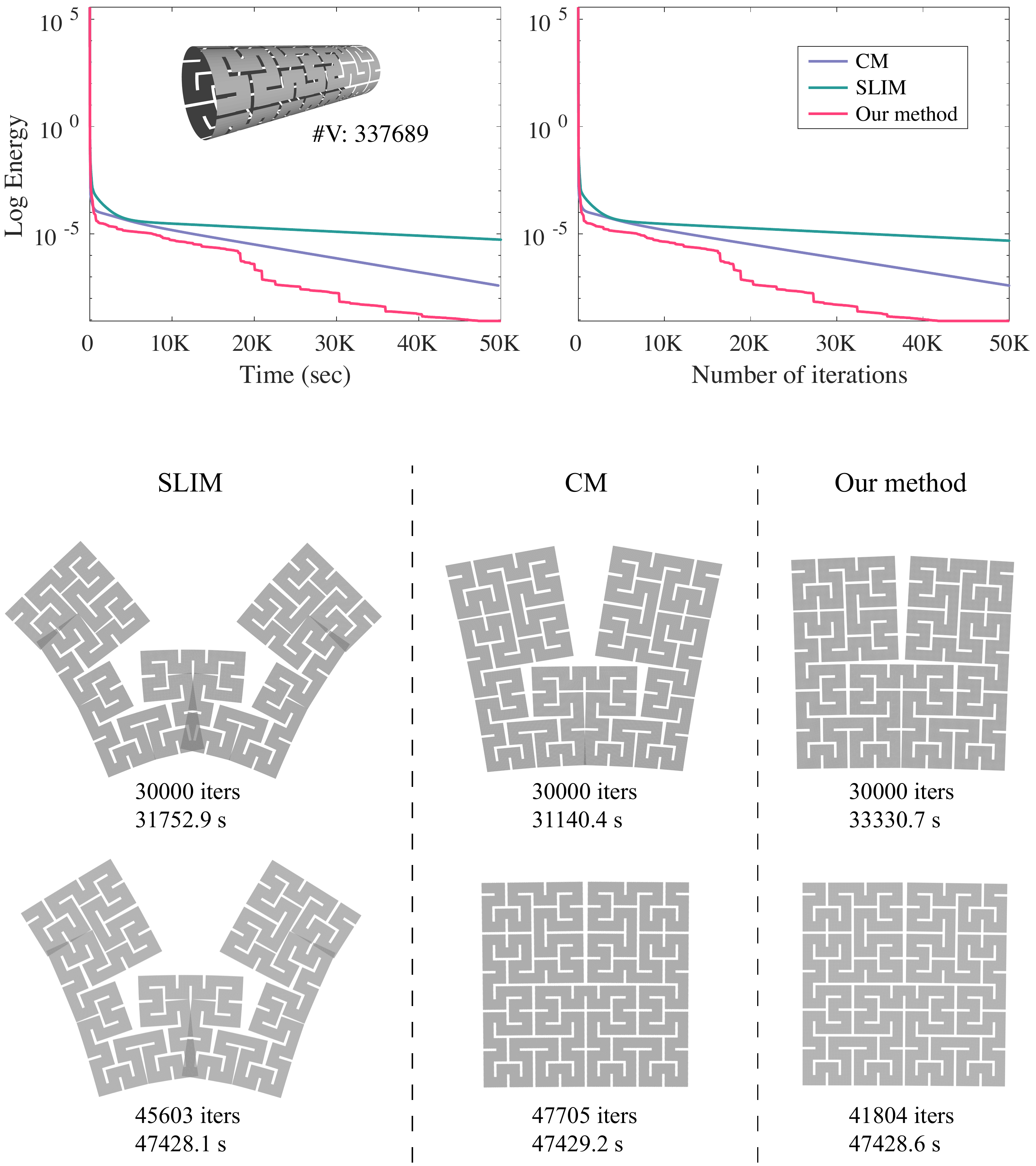}
	\caption{Unwrapping a Hilbert-curve-shape mesh on a cylinder by minimizing the symmetric Dirichlet energy.}
	\label{fig:HilbertTest}
\end{figure}

In Figure~\ref{fig:HilbertTest}, we compare our accelerated SLIM solver with the original SLIM and the CM solver, on a challenging problem proposed in~\cite{Smith2015}: we unwrap a Hilbert-curve-shaped mesh on a cylinder by optimizing the symmetric Dirichlet energy, starting from its Tutte's embedding. Since the mesh can be unwrapped without distortion, we evaluate the performance of each solver by plotting the difference between its resulting energy and the ground truth minimum energy. Our approach achieves the fastest decrease of energy and convergence to the solution.

\section{Discussion and Conclusion}

Although our acceleration is simple and effective on a variety of problems, there are a few limitations that we need to be aware of. First, although our analysis shows that the canonical mixing parameter $\beta = 1$ is effective for the local-global formulation, the same cannot be said for general fixed-point iteration solvers. Most existing work in the literature simply choose $\beta = 1$ and achieve good results. But it has also been shown that for problems, choosing another value can lead to better convergence~\cite{Fang2009}. It remains an open research problem to choose $\beta$ automatically for a general fixed-point solver. In the future, we would like to investigate the choice of $\beta$ beyond classical local-global solvers.

Our default selection strategy only compares the energy values of the accelerated iterate and the previous iterate. In theory, without comparison to the local-global iterate, it may accept an accelerated iterate that actually has a higher energy than its local-global counterpart and slows down the convergence. 
Moreover, although this strategy ensures monotonic decrease and convergence of the energy, it does not necessarily converge to a critical value of the energy, nor does it guarantee the convergence of variables.
So far we have not observed such pathological cases in our experiments. 
In the future, a more thorough investigation into the selection strategy and its convergence behavior would be helpful.

Fang and Saad~\shortcite{Fang2009} point out that Anderson acceleration is a particular case in the \emph{Anderson family} of multisecant methods, and call it the \emph{Type-II} method. Another member of the family, called the \emph{Type-I} method, approximates the Jacobian instead of its inverse, resulting in a slightly different formula for the accelerated iterate~\cite{Walker2011}. Most existing works, such as the local convergence proofs in~\cite{Toth2015,Toth2017}, are focused on the Type-II method. An interesting future work is to investigate the application and performance of the Type-I method.

The local-global solvers discussed in this paper enforce geometric constraints by minimizing their violation. Apart from fixed node positions, our approach currently does not support general hard constraints (i.e., constraints that need to be strictly satisfied). For geometry optimization and physics simulation, such hard constraints can be handled by introducing dual variables and using an augmented Lagrangian or ADMM solver~\cite{Deng2015,Overby2017}. In the future, we would like to extend our acceleration technique to solvers with both primal and dual variables.

To conclude, we propose in this paper a simple and effective way to apply Anderson acceleration to local-global solvers for geometry optimization and physics simulation. We improve upon the classical Anderson acceleration, by introducing a simple selection strategy that guarantees its global convergence with a small computational cost. In addition, we analyze and show the effectiveness of the canonical mixing parameter for achieving good acceleration results. 
Extensive experiments show that our technique achieves comparable or better results than state-of-the-art fast solvers on a variety of problems, and is applicable beyond the classical local-global solvers. 
Given the success of Anderson acceleration in other fields such as computational chemistry and computational physics, we believe it is a promising tool for improving existing algorithms and designing new algorithms for computer graphics.

\begin{acks}
	The initial PQ meshes in Figures~\ref{Fig:Planarity}, \ref{Fig:strategy} and \ref{Fig:ChoiceM} are provided by Yang Liu. The target model in Figure~\ref{Fig:wiremesh}, \href{https://www.thingiverse.com/thing:146386}{``Male Torso, Diadumenus Type''} by \href{https://www.thingiverse.com/CosmoWenman/about}{Cosmo Wenman}, is licensed under \href{https://creativecommons.org/licenses/by/3.0/}{CC BY 3.0}. This work was supported by the National Key R\&D Program of China (No. 2016YFC0800501), the National Natural Science Foundation of China (No. 61672481, No. 61672482 and No. 11626253), and the One Hundred Talent Project of the Chinese Academy of Sciences.
\end{acks}

\bibliographystyle{ACM-Reference-Format}
\bibliography{AAOptimization}
\end{document}